\documentclass[aps,prd,nofootinbib,preprint]{revtex4}

\usepackage{graphicx}% Include figure files
\usepackage{times}
\usepackage{multirow}
\usepackage{subfigure}
\usepackage{hyperref}
\usepackage{breakurl}
\usepackage{amsmath}
\usepackage[displaymath, mathlines]{lineno}
\usepackage{slashed}

\newcommand{\pslash}{\slashed{p}}
\newcommand{\gev}{\,{\rm GeV}}

\begin{document}

%\linenumbers

%%%%%%%%%%%%%%%%%%%%%%%%%%%%%%%%%%%%%%%%%%%%%%%%%%%%%%%%%%%%%%%%%%%%%%%%
% 
\title{Accessing proton generalized parton distributions and pion
distribution amplitudes with the exclusive pion-induced
Drell-Yan process at J-PARC}
\author{Takahiro Sawada,}
\email{sawada@phys.sinica.edu.tw}
\affiliation{Institute of Physics, Academia Sinica, Taipei 11529,
  Taiwan}

\author{Wen-Chen Chang,}
\email{changwc@phys.sinica.edu.tw}
\affiliation{Institute of Physics, Academia Sinica, Taipei 11529,
  Taiwan}

\author{Shunzo Kumano,}
\email{shunzo.kumano@kek.jp}
\affiliation{KEK Theory Center, Institute of Particle and Nuclear
  Studies, High Energy Accelerator Research Organization (KEK), 1-1,
  Oho, Tsukuba, Ibaraki 305-0801, Japan}
\affiliation{J-PARC Branch, KEK Theory Center, Institute of
  Particle and Nuclear Studies, KEK, 203-1, Shirakata, Tokai, Ibaraki
  319-1106, Japan}

\author{Jen-Chieh Peng,}
\email{jcpeng@illinois.edu}
\affiliation{Department of Physics, University of Illinois at
  Urbana-Champaign, Urbana, Illinois 61801, USA}

\author{Shinya Sawada,}
\email{shinya.sawada@kek.jp}
\affiliation{High Energy Accelerator Research Organization (KEK),
  1-1 Oho, Tsukuba, Ibaraki 305-0801, Japan}

\author{and Kazuhiro Tanaka}
\email{kztanaka@juntendo.ac.jp}
\affiliation{Department of Physics, Juntendo University, Inzai,
  Chiba 270-1695, Japan}
\affiliation{ J-PARC Branch, KEK Theory Center, Institute of
  Particle and Nuclear Studies, KEK, 203-1, Shirakata, Tokai, Ibaraki
  319-1106, Japan}


\begin{abstract}
Generalized parton distributions (GPDs) encoding multidimensional
information of hadron partonic structure appear as the building blocks
in a factorized description of hard exclusive reactions. The nucleon
GPDs have been accessed by deeply virtual Compton scattering and
deeply virtual meson production with lepton beam. A complementary
probe with hadron beam is the exclusive pion-induced Drell-Yan
process. In this paper, we discuss recent theoretical advances on
describing this process in terms of nucleon GPDs and pion distribution
amplitudes. Furthermore, we address the feasibility of measuring the
exclusive pion-induced Drell-Yan process $\pi^- p \to \mu^+\mu^- n$
via a spectrometer at the High Momentum Beamline being constructed at
J-PARC in Japan.  Realization of such measurement at J-PARC will
provide a new test of perturbative QCD descriptions of a novel class
of hard exclusive reactions. It will also offer the possibility of
experimentally accessing nucleon GPDs at large timelike virtuality.
\end{abstract}

%end of abstract

%\keywords{exclusive Drell-Yan process, timelike process, generalized
%  parton distribution, distribution amplitude}

%%%%%%%%%%%%%%%%%%%%%%%%%%%%%%%%%%%%%%%%%%%%%%%%%%%%%%%%%%%%%%%%%%%%%%%%
\maketitle

%%%%%%%%%%%%%%%%%%%%%%%%%%%%%%%%%%%%%%
\section{Introduction}
\label{sec:intro}
%%%%%%%%%%%%%%%%%%%%%%%%%%%%%%%%%%%%%%

Observation of the Bjorken-$x$ scaling behavior in charged-lepton deep
inelastic scattering clearly revealed quarks as the pointlike
constituents of nucleons~\cite{friedman}. In the Bjorken scaling
limit, structure functions of the nucleon are described in terms of
parton distribution functions (PDFs) only as a function of the scaling
variable $x$, which coincides with the parton's longitudinal momentum
fraction. The partonic number [$q(x)$] and helicity [$\Delta q(x)$]
distributions of nucleons have been well determined by global analysis
of extensive data mainly from the deep inelastic scattering and
Drell-Yan (DY) processes. The universality of nucleon PDFs extracted
from various high-energy scattering processes over wide kinematic
regions is a great success of perturbative QCD and factorization
theorems~\cite{PDF_Roeck2011,PDF_Perez2013,PDF_Forte2013,PDF_Delgado2013,
  polPDF_Lampe1998,polPDF_deFlorian2011,polPDF_Aidala2012}.

Current experimental information indicates that quark- and
gluon-helicity contributions cannot fully account for the nucleon
spin. The orbital-angular-momentum contribution could provide the
missing piece for the nucleon spin. In recent decades, tremendous
efforts have been spent in extending the measurement of partonic
structure of nucleons to multidimension: generalized parton
distributions (GPDs)~\cite{Goeke:2001tz,Diehl:2003ny,Ji:2004gf,
  Belitsky:2005qn,Boffi:2007yc,Diehl:2013xca} and
transverse-momentum-dependent parton distribution functions
(TMDs)~\cite{polPDF_Aidala2012,D'Alesio:2007jt,Barone:2010zz,Perdekamp:2015vwa}.
The multidimensional information becomes essential for a deeper
understanding of the partonic structures of the nucleon, including the
origin of the nucleon spin.

\begin{figure}[htbp]
\begin{center}
 \includegraphics[width=1.0\textwidth]{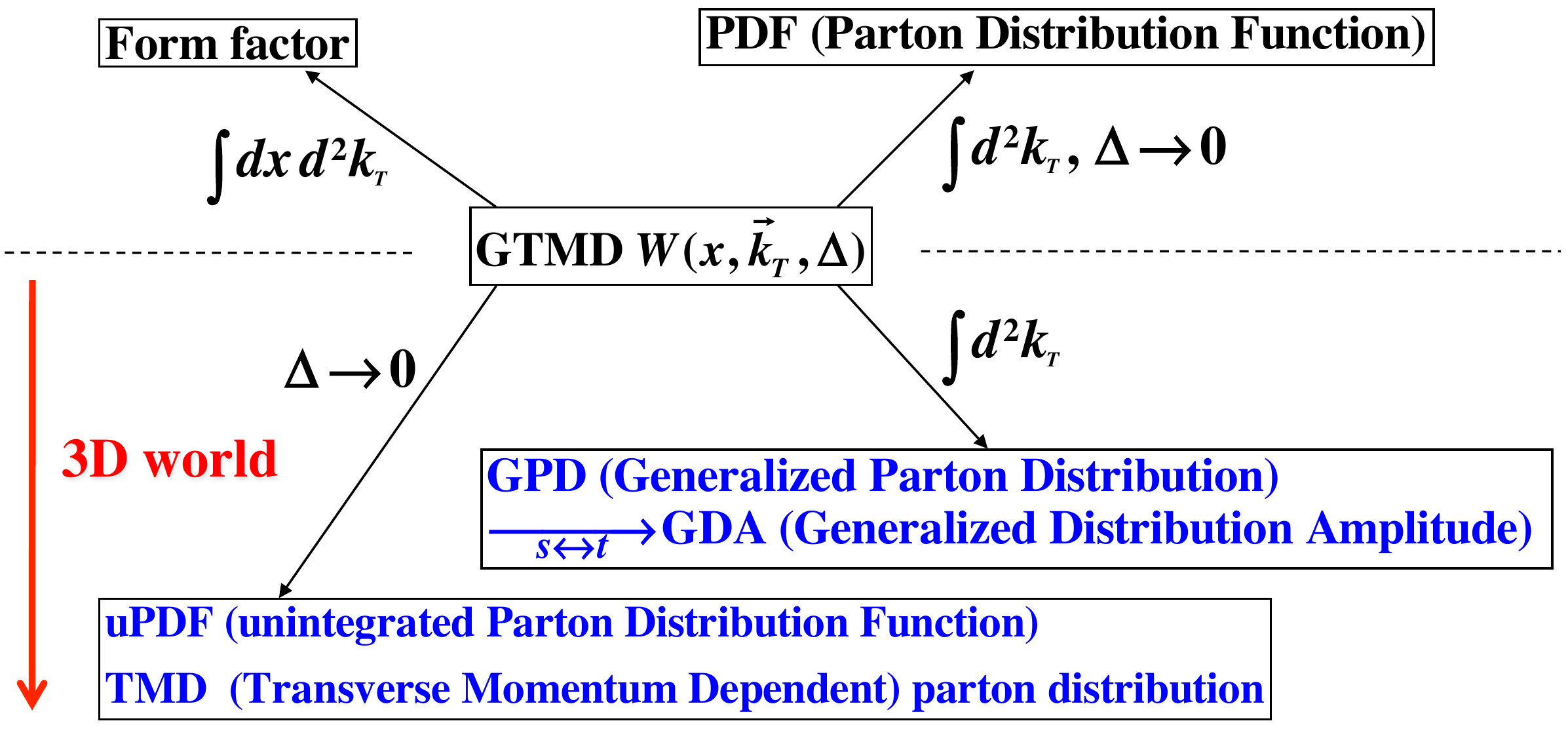}
\end{center}
%\vspace{-1.5cm}
\caption{Three-dimensional structure functions from the GTMD.  Here,
  $\vec k_T$ is the parton's transverse momentum and $\Delta$
  indicates the momentum transfer for the off-forward matrix element.
}
\label{fig:3d-structure}
\end{figure}

In addition to the longitudinal momentum fraction $x$, the GPDs encode
the dependence on transverse spatial distributions while TMDs include
that of intrinsic transverse momentum $(k_T$) of partons. The TMDs are
also called unintegrated PDFs in the
unpolarized case, and the $s$-$t$ crossed quantities of the GPDs are
generalized distribution amplitudes. These distributions are
obtained from the generating functions, the generalized
transverse-momentum-dependent parton distribution functions
(GTMDs)~\cite{gtmd}, by integrating over some kinematical variables
or/and by taking the forward ($\Delta =0$) limit as shown in
Fig.~\ref{fig:3d-structure}. The GTMDs depend on the partonic
variables, the longitudinal momentum fraction $x$ and the transverse
momentum $\vec k_T$, as well as on the momentum transfer $\Delta$
associated with the off-forward matrix element. We will focus on the
study of nucleon GPDs in this work.

In Fig.~\ref{fig:3d-structure}, we showed various functions derived
from the GTMD. We note that the GTMD has a direct connection to
``Wigner
distribution''~\cite{Wigner,Belitsky:2005qn,lp-2011,lplp}. The
original definition of the Wigner distribution in Ref.~\cite{Wigner}
intended to provide the six-dimensional phase-space distribution, $W (x,
\vec k_T, \vec r \,)$, corresponding to the generating function of the
other various distributions. However, it was formulated in a
particular Lorentz frame and appeared to be plagued by relativistic
corrections. Another definition of the Wigner distribution is
given~\cite{lp-2011} in the infinite momentum frame, providing the
five-dimensional phase-space distribution, $W (x, \vec k_T, \vec r_T
\,)$, which is now known to correspond to the $\Delta^+ = 0$ ($\xi =
0$) limit of the GTMD~\cite{lplp}.

Motivated by the orbital-angular-momentum contribution of partons to
the nucleon spin, GPDs~\cite{GPD} were introduced in connection with
two hard exclusive processes of leptoproduction of photons and mesons
off protons: deeply virtual Compton scattering
(DVCS)~\cite{Ji:1996ek,Radyushkin:1996nd,Ji:1996nm} and deeply virtual
meson production (DVMP)~\cite{Collins:1996fb,Collins:1998be}. With a
factorization of perturbatively calculable short-distance hard part
and universal long-distance soft hadronic matrix elements, the nucleon
GPDs, which are the common soft objects, could be obtained from the
measurement of these two processes.

There have been tremendous experimental efforts on measuring DVCS and
DVMP processes with electron beam. Data have been taken by HERMES, H1
and ZEUS at DESY and HALL-A and CLAS at JLab. Recently the status of
``global analysis'' of nucleon GPDs in the valence region with
existing DVCS and DVMP data is reviewed in Refs.~\cite{Guidal:2013rya}
and ~\cite{Favart:2015umi} respectively. Further
experiments~\cite{Guidal:2013rya} are planned for JLab after the 12-GeV
upgrade~\cite{Dudek:2012vr} and the COMPASS experiment at CERN with muon
beam~\cite{COMPASSII}.

Other than lepton beams, it was suggested that GPDs could also be
accessed using real photon and hadron beams, such as timelike Compton
scattering~\cite{Berger:2001xd}, lepton-pair production with meson
beam~\cite{Berger:2001zn,Goloskokov:2015zsa}, and pure hadronic
reaction~\cite{Kumano:2009he,Kawamura:2013wfa}. For example, invoking
the properties under time-reversal transformation and analyticity
under the change from spacelike to timelike large virtuality, the
exclusive pion-induced Drell-Yan process $\pi N \to \gamma^*
N$~\cite{Berger:2001zn,Goloskokov:2015zsa} is assumed to obey a
factorization similar to the DVMP processes and can serve as a
complementary timelike probe to access nucleon
GPDs~\cite{Muller:2012yq}. Such a measurement is interesting as well
as important to verify the universality of GPDs in both spacelike and
timelike processes. There is a unique opportunity to carry out the
measurements of the exclusive Drell-Yan process using the
high-intensity hadron beams at the Japan Proton Accelerator Research
Complex (J-PARC). A related feasibility study of accessing
pion-nucleon transition distribution amplitudes from $\bar{p} p \to
\gamma^* \pi$ with the $\bar{\rm P}$ANDA experiment at FAIR was done in
Ref.~\cite{Singh:2014pfv}.

The present paper is aimed at the feasibility study of measuring the
exclusive pion-induced Drell-Yan process in the upcoming high-momentum
beam line of J-PARC. It is organized as follows. In
Sec.~\ref{sec:tomography}, we briefly introduce nucleon GPDs and pion
distribution amplitudes (DAs), two nonperturbative partonic structures
to be extracted. The theoretical formalism of exclusive pion-induced
Drell-Yan process is discussed, and the predicted differential cross
sections are given in Sec.~\ref{sec:exc_dy}. We then address the
possibility of detecting the exclusive pion-induced DY events with the
spectrometer of the E50 experiment at J-PARC Hadron Hall. We conclude
the paper in Sec.~\ref{sec:summary}.

%%%%%%%%%%%%%%%%%%%%%%%%%%%%%%%%%%%%%%
\section{Hadron tomography}
\label{sec:tomography}
%%%%%%%%%%%%%%%%%%%%%%%%%%%%%%%%%%%%%%

In the leading-order handbag approach for the process of deeply
virtual pion production, say $\gamma^* p \to \pi N$, the amplitudes
could be factorized into a perturbative hard part and universal soft
hadronic matrix elements. The latter part is composed of two objects
parametrized as proton GPDs and pion DAs. For these hard exclusive
processes, factorization was proven up to the leading twist-two level
in the collinear
framework~\cite{Collins:1996fb,Collins:1998be}. Swapping the initial
and final states, and replacing the momentum of $\gamma^*$ with the
timelike one, the factorization formalism is proposed to be applicable
to the exclusive Drell-Yan process, $\pi N \to \gamma^* N$, with the
same universal nonperturbative
input~\cite{Berger:2001zn,Muller:2012yq}. Therefore, in the exclusive
pion-induced Drell-Yan process, to be introduced in details later, the
nucleon GPDs and pion DAs are both present in the factorization
formulation of the cross sections. Below we give a brief introduction
on these two soft objects.

\subsection{Nucleon generalized parton distributions}
\label{sec:gpds}

\begin{figure}[htbp]
\begin{center}
   \includegraphics[width=0.5\textwidth]{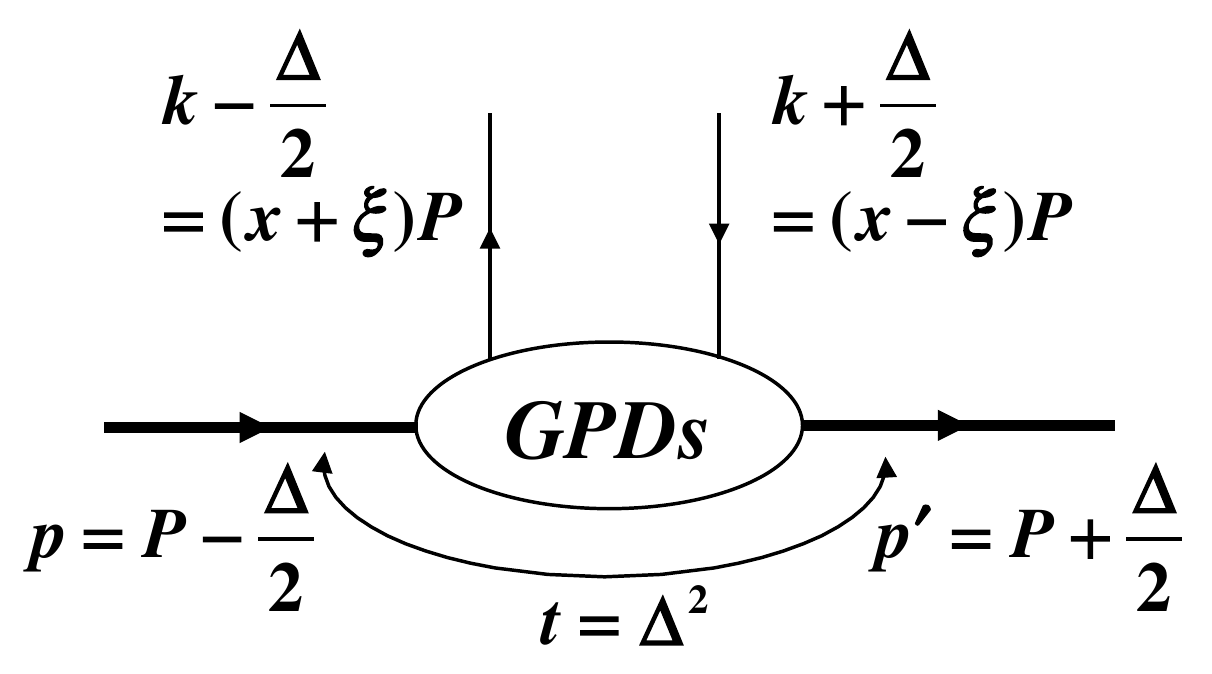}
\end{center}
\caption{Kinematics for GPDs.}
\label{fig:gpd-kinematics}
\end{figure}

Among the three-dimensional structure functions, the GPDs have a clear
connection with the orbital-angular-momentum contribution to the
nucleon spin. The kinematics for the GPDs is shown in
Fig. \ref{fig:gpd-kinematics}, where $p$ and $p\,'$ are initial and
final nucleon momenta, respectively.  The corresponding average
momentum and momentum transfer are denoted as $P$ $(\,=(p+p\,')/2\,)$
and $\Delta$ $(\, =p\,' -p\,)$, respectively, and the momentum
transfer squared is defined as $t=\Delta^2 =(p\,' -p)^2$. The momenta
of outgoing and incoming quarks are denoted as $k-\Delta/2$ and
$k+\Delta/2$ with the average momentum $k$.

In a reference frame with the average nucleon momentum $P$ pointing
along the positive \^{z} axis, the scaling variable $x$ and a skewness
parameter $\xi$ are defined as
\begin{equation}
x = \frac{\left( (k-\Delta/2)+(k+\Delta/2) \right)^+}{(p+p\,')^+}=\frac{k^+}{P^+}
\label{eqn:x}
\end{equation}
and
\begin{equation}
\xi= \frac{(p-p\,')^+}{(p+p\,')^+} = \frac{-\Delta^+}{2P^+} =
\frac{\left( (k-\Delta/2)-(k+\Delta/2) \right)^+}{(p+p\,')^+},
\label{eqn:xi}
\end{equation}
respectively, where $a^\pm=(a^0 \pm a^3)/\sqrt{2}$ denote the
plus/minus light-cone components of a four-vector $a^\mu$.  Here, $x$
and $2\xi$ are the light-cone momentum fractions of the average
momentum and momentum transfer for the relevant quarks, respectively,
to the average momentum of the parent nucleon. The range of $x$ is
from $-1$ to $1$ while $\xi$ is between $0$ and $1$.

The GPDs for the nucleon are defined by off-forward nucleon matrix
elements of quark (and gluon) bilocal operators with a lightlike
separation. The quark GPDs relevant to the processes without the
quark-helicity flip are given by (see,
e.g.,~\cite{Diehl:2003ny,Belitsky:2005qn})
\begin{align}
 \int & \frac{d y^-}{4\pi}e^{i x P^+ y^-}
 \left< p' \left| 
 \bar{q}(-y/2) \gamma^+ q(y/2) 
 \right| p \right> \Big |_{y^+ = \vec y_\perp =0}
\nonumber \\
 = & \frac{1}{2  P^+} \, \bar{u} (p') 
 \left [ H^q (x,\xi,t) \gamma^+
     + E^q (x,\xi,t)  \frac{i \sigma^{+ \alpha} \Delta_\alpha}{2 \, m_N}
 \right ] u (p) ,
\label{eqn:gpd-n}
\end{align}
and
\begin{align}
 \int & \frac{d y^-}{4\pi}e^{i x P^+ y^-}
 \left< p' \left| 
 \bar{q}(-y/2) \gamma^+ \gamma_5 q(y/2) 
 \right| p \right> \Big |_{y^+ = \vec y_\perp =0}
\nonumber \\
 = & \frac{1}{2  P^+} \, \bar{u} (p') 
 \left [ \tilde{H}^q (x,\xi,t) \gamma^+ \gamma_5
     + \tilde{E}^q (x,\xi,t)  \frac{\gamma_5 \Delta^+}{2 \, m_N}
 \right ] u (p) ,
\label{eqn:gpd-p}
\end{align}
for each quark flavor $q$, where $ \left| p \right>$ denotes the
proton state with momentum $p$ and mass $m_N$, $u(p)$ denote the Dirac
spinor for the proton, and $\sigma^{\alpha\beta}$ is given by
$\sigma^{\alpha\beta}=(i/2)[\gamma^\alpha, \gamma^\beta]$.  Here and
below, for simplicity, we do not show the gauge-link operator between
the two quark fields for maintaining the gauge invariance.  $H^q
(x,\xi,t)$ and $E^q (x,\xi,t)$ are the unpolarized quark GPDs, and
$\tilde{H}^q (x,\xi,t)$ and $\tilde{E}^q (x,\xi,t)$ are the polarized
ones. We have also suppressed the renormalization scale dependence of
these GPDs originating from that of the bilocal operator in
Eqs.~(\ref{eqn:gpd-n}) and (\ref{eqn:gpd-p}).

We recall three important features of the GPDs.
\begin{enumerate}
\itemsep0em

\item[(1)] The $H^q (x,\xi,t)$ and $\tilde{H}^q (x,\xi,t)$ GPDs become
  the unpolarized and helicity PDFs for the nucleon in the forward
  limit ($\Delta\to 0$, $\xi\to 0$ and $t \rightarrow 0$):
\begin{align*}
H^q (x, 0, 0) &= q(x), & \tilde{H}^q (x, 0, 0) &= \Delta q(x).
\end{align*}

\item[(2)] The first moments of $H^q (x,\xi,t)$ and $E^q (x,\xi,t)$
  are Dirac and Pauli form factors of the nucleon and those of
  $\tilde{H}^q (x,\xi,t)$ and $\tilde{E}^q (x,\xi,t)$ are axial and
  pseudoscalar form factors:
\begin{align}
\int_{-1}^{1} dx H^q(x,\xi,t)  & = F_1^q (t), & 
\int_{-1}^{1} dx E^q(x,\xi,t)  & = F_2^q (t),\nonumber\\
\int_{-1}^{1} dx \tilde{H}^q(x,\xi,t)  & = g_A^q (t), &
\int_{-1}^{1} dx \tilde{E}^q(x,\xi,t)  & = g_P^q (t), 
\label{sumrule}
\end{align}
for each separate quark flavor. The $x^n$ moments of GPDs are
polynomials in $\xi$ of order $n+1$ or $n$~\cite{Diehl:2003ny}. This
``polynomiality'' property is important in constraining the $x$ and
$\xi$ dependence of GPDs.

\item[(3)] The second $x$ moments are also related to matrix elements
  of certain local operators, in particular, to matrix elements of the
  quark angular-momentum operator: $J^q = \int_{-1}^{1} dx \, x \, [
    H^q (x,\xi,t=0) +E^q (x,\xi,t=0) ] / 2$, which allows us to deduce
  the quark orbital-angular-momentum contribution ($L^q$) to the
  nucleon spin as $J^q = \Delta q /2 + L^q$. The absence of $\xi$
  dependence in the sum is due to the cancellation of individual
  $\xi$-dependent terms of $H^q$ and $E^q$.
\end{enumerate}
Thus, the GPDs are considered to be the key quantities to resolve the
three-dimensional structure of the nucleon, including the
long-standing issue of nucleon-spin origin in terms of the
orbital-angular-momentum contribution.

\begin{figure}[htbp]
\begin{center}
\includegraphics[width=1.0\textwidth]{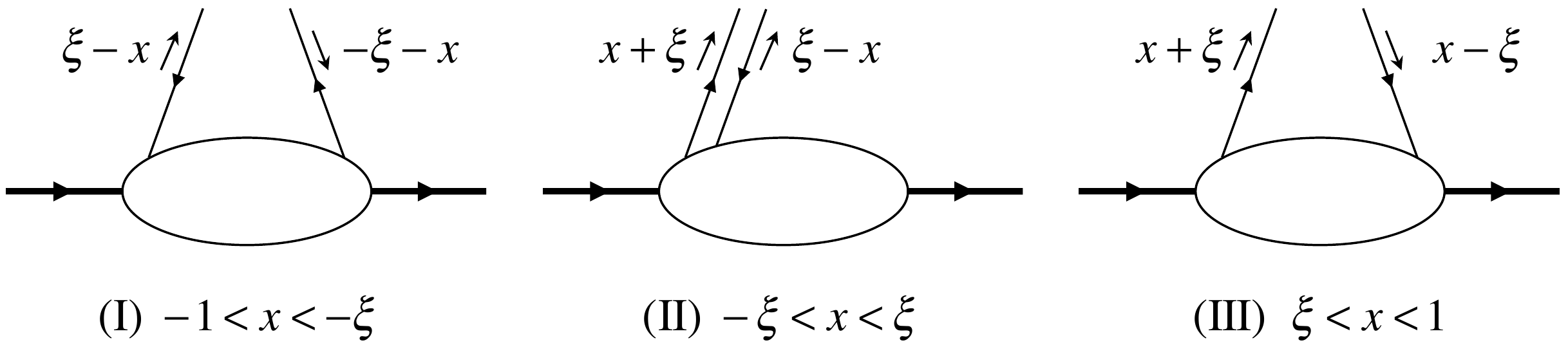}
\end{center}
\caption{Three $x$ regions of GPDs: (a) [-1, -$\xi$], (b) [-$\xi$, $\xi$] and
  (c) [$\xi$, 1]. The short arrows point the direction of parton's
  momentum.}
\label{fig:gpd-regions}
\end{figure}

There are three kinematical regions for the GPDs as shown in
Fig.~\ref{fig:gpd-regions}. They correspond to the following three
kinds of distributions.
\begin{enumerate}
\itemsep0em

\item[($\rm{I}$)] ``Antiquark distribution" at $-1 < x < -\xi$ \ ($x+\xi
  <0$ and $x-\xi <0$): emission of an antiquark with momentum fraction
  $\xi -x$ and absorption of an antiquark with momentum fraction
  $-x-\xi$.

\item[($\rm{II}$)] ``Quark-antiquark distribution amplitude" at $-\xi < x
  <\xi$ \ ($x+\xi >0$ and $x-\xi <0$): emission of a quark with
  momentum fraction $x+\xi$ and an antiquark with momentum fraction
  $\xi-x$.

\item[($\rm{III}$)] ``Quark distribution" at $\xi < x <1$ \ ($x+\xi >0$ and
  $x-\xi >0$): emission of a quark with momentum fraction $x+\xi$ and
  absorption of a quark with momentum fraction $x-\xi$.
\end{enumerate}

The regions ($\rm{I}$) and ($\rm{III}$), called the DGLAP
(Dokshitzer-Gribov-Lipatov-Altarelli-Parisi) regions, and ($\rm{II}$),
called the ERBL (Efremov-Radyushkin-Brodsky-Lepage) region, have
different types of corresponding evolution equations. For clarifying
the three-dimensional structure of the nucleon, including the origin
of the nucleon spin, all of these regions should be investigated.

In principle, all four GPDs, $H^q (x,\xi,t)$, $E^q (x,\xi,t)$,
$\tilde{H}^q (x,\xi,t)$ and $\tilde{E}^q (x,\xi,t)$, contribute to the
spacelike processes like DVCS and DVMP as well as timelike ones like
hard exclusive hadronic reactions and exclusive Drell-Yan. Precise
determination of them requires global analyses of measurements
covering a broad kinematic range with lepton as well as hadron
beams. For the processes associated with hadron beam and/or hadron
production, the initial proton may change into the neutron in the
final state, and we need the transition GPDs, which are a
straightforward extension of Eqs.~(\ref{eqn:gpd-n}) and
(\ref{eqn:gpd-p}), e.g.,

\begin{align}
 &\int \frac{d y^-}{4\pi}e^{i x P^+ y^-}
 \left< n(p') \left| 
 \bar{d}(-y/2) \gamma^+ \gamma_5 u(y/2) 
 \right| p(p) \right> \Big |_{y^+ = \vec y_\perp =0}
\nonumber \\
 &= \frac{1}{2  P^+} \, \bar{u} (p') 
 \left [ \tilde{H}^{du} (x,\xi,t) \gamma^+ \gamma_5
     + \tilde{E}^{du} (x,\xi,t)  \frac{\gamma_5 \Delta^+}{2 \, m_N}
 \right ] u (p) .
\label{eqn:gpd-ppn}
\end{align}
Using isospin invariance, these transition GPDs can be expressed in
terms of the combination of proton GPDs of Eq.~(\ref{eqn:gpd-p}) with
$q=u,d$, as~\cite{Mankiewicz:1997aa}
\begin{align}
\tilde{H}^{du} (x,\xi,t) &= \tilde{H}^u (x,\xi,t)-\tilde{H}^d (x,\xi,t), \nonumber \\
\tilde{E}^{du} (x,\xi,t) &= \tilde{E}^u (x,\xi,t)-\tilde{E}^d (x,\xi,t).
\label{isospinrelation}
\end{align}

\noindent
There exist other transition GPDs with the final state as $\Delta$ or
a $N^*$ resonance. Although there are some theoretical studies on
these transition GPDs~\cite{Goeke:2001tz,Kumano:2009he,transition-gpds}, they have not been
measured yet. Their effects on the measurement of nucleon GPDs are not
taken into account in the current study and require further work.

\subsection{Pion distribution amplitude}
\label{sec:pion}

The pion DAs are defined as the pion-to-vacuum matrix element of a
bilocal quark operator as in Eqs.~(\ref{eqn:gpd-p}) and
(\ref{eqn:gpd-ppn}). For a $\pi^-$ meson with the momentum $p_\pi$
pointing along the negative \^{z} axis~\cite{erbl,pionDA:CZ},
\begin{align}
\langle \, 0 \, | \, \bar u (0)_\alpha \, 
                  \, d (y)_\beta \,
                | \, \pi^- (p_\pi) \, \rangle\Big |_{y^- = \vec y_\perp =0}
= \frac{i f_\pi}{4} \int_0^1 dx \, e^{-i x p_\pi^- y^+}
  \left ( \gamma_5 \, \pslash_\pi \right )_{\beta\alpha} \, \Phi_\pi (x,\mu) + \cdots,
\label{eqn:qbar-q-matrix}
\end{align}
where $\Phi_\pi (x,\mu)$ denotes the twist-two DA and the ellipses
stand for the higher-twist terms. The factor $f_\pi$ is the pion decay
constant defined as $\langle \, 0 \, | \, \bar u (0) \gamma^\mu
\gamma_5 \, d(0) \, | \, \pi^- (p_\pi) \, \rangle = i f_\pi
p_\pi^\mu$, such that the DA obeys the normalization condition,
\begin{equation}
\int_0^1 dx \, \Phi_\pi (x, \mu)=1.
\label{normcon}
\end{equation}
Here, $\Phi_\pi (x,\mu)$ is expressed as a function of the two
variables $x$ and $\mu$: $x$ is the longitudinal momentum fraction of
a valence quark in the pion and $\mu$ is the renormalization scale of
the bilocal operator in Eq.~(\ref{eqn:qbar-q-matrix}). The scale
($\mu$) dependence of $\Phi_\pi (x,\mu)$ is described by the ERBL-type
evolution equation~\cite{erbl}.

Using translational invariance and changing the integration variable
as $x \rightarrow (z+1)/2$, the definition (\ref{eqn:qbar-q-matrix})
can be recast as
\begin{align}
\langle \, 0 \, | \, \bar u (-y)_\alpha \, 
                  \, d (y)_\beta \,
                | \, \pi^- (p_\pi) \, \rangle\Big |_{y^- = \vec y_\perp =0}
= \frac{i f_\pi}{4} \int_{-1}^1 dz \, e^{-i z p_\pi^- y^+}
  \left ( \gamma_5 \, \pslash_\pi \right )_{\beta\alpha} \, \phi_\pi (z,\mu) + \cdots,
\label{eqn:qbar-q-matrixnew}
\end{align}
with 
\begin{equation}
\phi_\pi (z,\mu)=\frac{1}{2}\Phi_\pi\left(x= \frac{z+1}{2},
\mu\right) \;\;\;\; \mathrm{and} \;\;\;\; \int_{-1}^1 dz \, \phi_\pi
(z, \mu)=1.
\label{normconnew}
\end{equation}

In the light-cone quantization formalism, a pion state can be expanded
by the Fock states constructed with physical degrees of freedom of
quarks and gluons as
\begin{equation}
|\pi(p_\pi)\rangle=\int \frac{dx}{\sqrt{x\bar{x}}}
\frac{d^2 \vec k_T}{16\pi^3}
\Psi_{q\bar{q}/\pi}(x, \vec k_T)
|q(k_q)\bar{q}(k_{\bar{q}})\rangle+\cdots ,
\label{eqn:BS}
\end{equation} 
where $|q(k_q)\bar{q}(k_{\bar{q}})\rangle$ is the leading $q\bar{q}$
Fock state and $\Psi_{q\bar{q}/\pi}(x, \vec k_T)$ denotes the
corresponding Bethe-Salpeter (BS) wave function, while the ellipses
stand for the contributions of higher Fock states, e.g., $|q\bar{q}
g\rangle$, $|q\bar{q} q\bar{q} \rangle$, etc. The $p_\pi \simeq
(0^+,p_\pi^-, \vec 0_T)$ is the pion momentum, and $\bar x$ is defined
by $\bar{x}=1-x$. The quark and antiquark in the leading Fock state
have the (off-shell) momenta with $k_q^-= xp_\pi^-$, $k_{\bar{q}}^-=
\bar{x}p^-_\pi$, and $\vec{k}_{q T}=-\vec{k}_{\bar{q} T}=\vec
k_T$. The normalization of the BS wave function is given by
$\int_0^1dx\int d^2 \vec k_T / (16\pi^3) |\Psi_{q\bar{q}/\pi}(x,\vec
k_T)|^2 = 1$, up to the additional positive terms from the higher Fock
states. The BS wave function is related to the light-cone DA
as~\cite{mueller-1989}
\begin{equation}
\int_{|\vec k_T|<\mu} \frac{d^2 \vec k_T}{16\pi^3}
\Psi_{d\bar{u}/\pi}(x,\vec k_T)
=
\frac{if_\pi}{4}\sqrt{\frac{2}{N_c}}\Phi_\pi(x,\mu) ,
\label{eqn:BS-3}   
\end{equation}
where $N_c$ is the number of colors.

The asymptotic form of the leading-twist distribution amplitude at the
formal limit $\mu \to \infty$ is known as~\cite{erbl}
\begin{equation}
\Phi_\pi^{\text{as}} (x) = 6 \, x \, (1-x) ,
\label{eqn:asymp}
\end{equation}
whereas it is generally expressed by the expansion in terms of
the Gegenbauer polynomials $C_n^{3/2} (2x-1)$ at finite $\mu$:
\begin{equation}
\Phi_\pi (x,\mu) = 6 \, x \, (1-x) 
\sum_{n=0,2,4,\cdots}^\infty a_n (\mu) \, C_n^{3/2} (2x-1).
\label{eqn:Gegenbauer}
\end{equation}
Here, the summation is taken over even numbers due to isospin
symmetry, which requires that DA is symmetric under $x \rightarrow
1-x$, i.e., $\Phi_\pi(1-x, \mu) = \Phi_\pi(x, \mu)$. 

Substituting the expansion (\ref{eqn:Gegenbauer}) into the definition
(\ref{eqn:qbar-q-matrix}) and using orthogonality relations of the
Gegenbauer polynomials, it is straightforward to see that $a_n (\mu)$
is given by matrix element of the local quark operator, whose
renormalization scale dependence determines the $\mu$ dependence of
$a_n(\mu)$, such that $a_n(\mu)$ is suppressed as $\mu$ increases and
the corresponding suppression is stronger for larger $n$. Moreover,
the rapidly oscillating behavior of the Gegenbauer polynomials of high
$n$ would lead to the suppression of the relevant convolution
integral. Therefore, in view of the present accuracy of experimental
data as well as theoretical calculations, $a_n$ with $n=4, 6, \cdots$,
in the expansion (\ref{eqn:Gegenbauer}) may be set to zero. 

\begin{table}[hbtp]
\begin{center}
\caption{Modeling of pion DAs}
\begin{tabular}{|c|c|c|c|c|}
  \hline
  \hline
  $\phi_{\pi}(z,\mu)$ & Asymptotic~\cite{erbl} & CZ~\cite{pionDA:CZ} & GK~\cite{Kroll:2012sm} & DSE~\cite{pionDA:DSE} \\
  \hline
  $a_2$ & 0 & 2/3 & 0.22 & 0.20 \\
  $a_4$ & 0 & 0 & 0 & 0.093 \\
  $a_6$ & 0 & 0 & 0 & 0.055 \\
  $\mu^2$ (GeV$^2$) & 1 & 0.25 & 4 & 4 \\
  \hline
  \hline
\end{tabular}
\label{tab:pionDA}
\end{center}
\end{table}

On the other hand, there are theoretical evidences that the first
coefficient $a_2$ is positive: lattice QCD~\cite{Braun:2006dg} and QCD
sum
rules~\cite{pionDA:CZ,Braun:1988qv,Ball:2006wn,Khodjamirian:1997tk,Braun:1999uj}
indicate that the $x$ distribution is broader than the asymptotic form
(\ref{eqn:asymp}). For example, based on the QCD sum rule
calculations, Chernyak and Zhitnitsky (CZ) proposed a form with
$a_2\,(\mu\simeq 0.5~{\rm GeV}) = 2/3$~\cite{pionDA:CZ},
\begin{align}
\Phi_\pi^{\text{CZ}} (x, \mu\simeq 0.5~{\rm GeV}) &= 30 \, x \, (1-x) \,
(2x-1)^2 \nonumber \\ 
&= 6 \, x \, (1-x) \, \left[ 1 + \frac{2}{3}
  C_2^{3/2} (2x-1) \right] .
\label{eqn:CZ}
\end{align}
A remarkable difference of this function from the asymptotic form is
that its value at $x=1/2$ is minimum and vanishes
($\Phi_\pi^{\text{CZ}}\,(x=0.5, \mu\simeq 0.5~{\rm GeV})=0$), whereas
it is maximum in the asymptotic form (\ref{eqn:asymp}).  There have
been other studies suggesting values of $a_2$ smaller than that of
Eq.~(\ref{eqn:CZ})~\cite{Braun:2006dg,Ball:2006wn,Khodjamirian:1997tk,Braun:1999uj}. In
particular, recent measurements for $\gamma \, \gamma^* \to \pi^0$ at
{\it BABAR}~\cite{Aubert:2009mc} and Belle~\cite{Uehara:2012ag} imposed a
constraint for the pion DA.  A value of $a_2\,(\mu=2~{\rm GeV})=0.22$
used in Ref.~\cite{Kroll:2012sm} is consistent with the {\it BABAR} data,
while the Belle data are compatible with a pion DA close to the
asymptotic form (\ref{eqn:asymp}). Also, the evaluation of the pion DA
using the Dyson-Schwinger equation (DSE) framework gave a recent
estimate of the Gegenbauer moments $a_4, a_6$, as well as
$a_2$~\cite{pionDA:DSE}.

\begin{figure}[htbp]
\begin{center}
\includegraphics[width=0.7\textwidth]{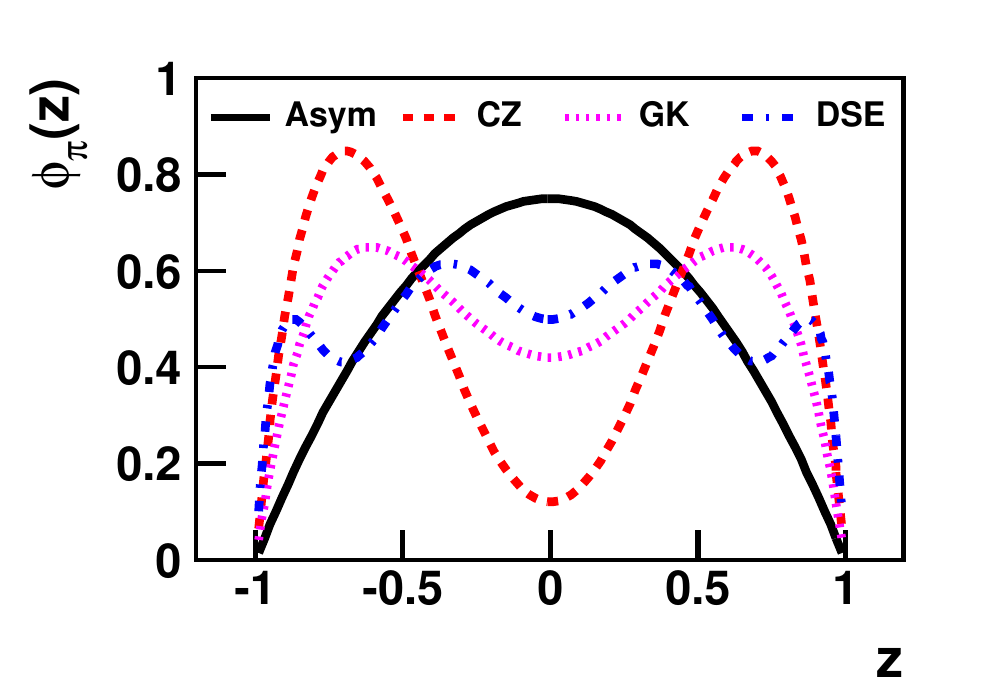}
\caption
[\protect{}] {Pion DAs [$\phi_\pi(z)$] at renormalization scale $\mu =
  1$ GeV: asymptotic form (solid)~\cite{erbl}, CZ model
  (dashed)~\cite{pionDA:CZ}, Goloskokov-Kroll (GK) model
  (dotted)~\cite{Kroll:2012sm} and DSE
  (dot-dashed)~\cite{pionDA:DSE}.}
\label{fig:pionDA}
\end{center}
\end{figure}

Table~\ref{tab:pionDA} lists the values of the relevant Gegenbauer
moments of different modeling for the pion DAs (\ref{eqn:Gegenbauer})
and their corresponding distributions using Eq.~(\ref{normconnew}) at
$\mu=1$~GeV are illustrated in Fig.~\ref{fig:pionDA}. Later we will
study the dependency of production cross sections for the exclusive
Drell-Yan process on the pion DAs.

%%%%%%%%%%%%%%%%%%%%%%%%%%%%%%%%%%%%%%
\section{Exclusive pion-induced Drell-Yan process}
\label{sec:exc_dy}
%%%%%%%%%%%%%%%%%%%%%%%%%%%%%%%%%%%%%%

\subsection{From semi-exclusive to exclusive}
\label{sec:exc_dy0}

\begin{figure}[htbp]
\begin{center}
\centering
\subfigure[]
{\includegraphics[width=0.48\textwidth]{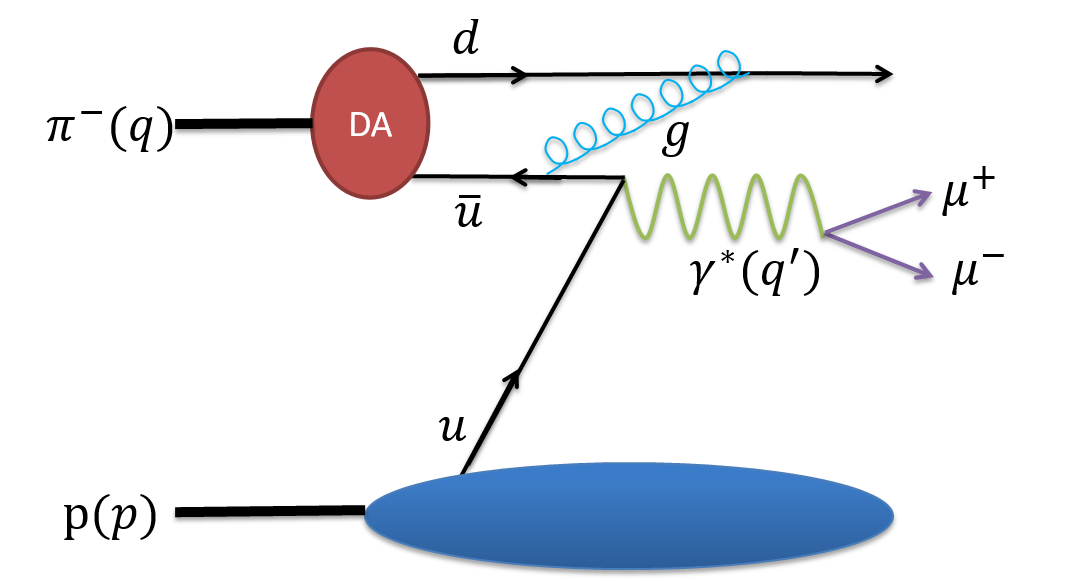}
\label{fig:edycross1}}
\subfigure[]
{\includegraphics[width=0.48\textwidth]{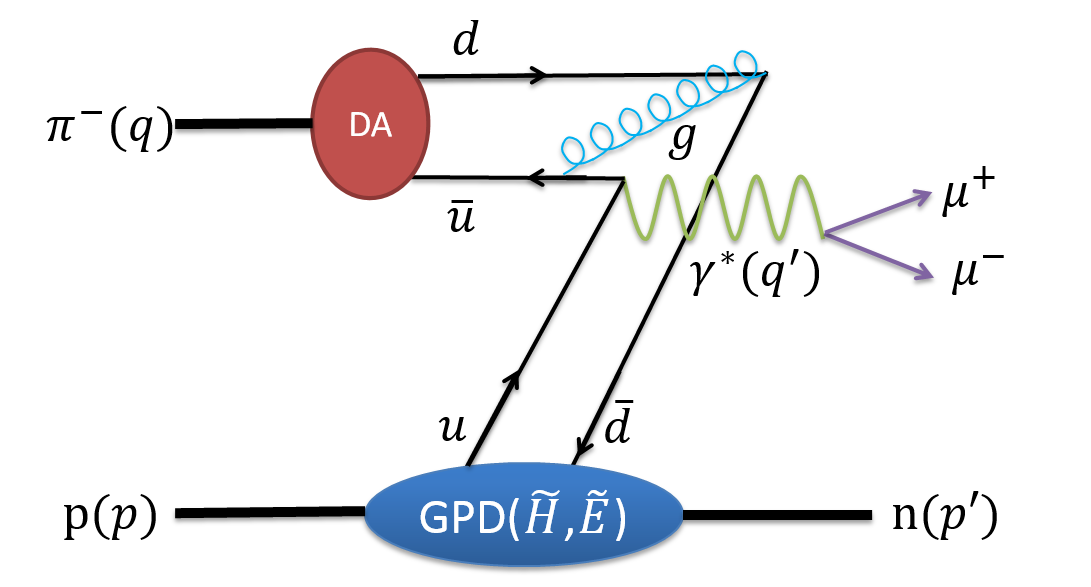}
\label{fig:edycross2}}
\caption
[\protect{}] {(a) Semi-exclusive pion-induced Drell-Yan process at
  large $x_{\pi}$ (b) Exclusive pion-induced Drell-Yan process.}
\label{fig:edycross}
\end{center}
\end{figure}

It has been experimentally observed that the polarization direction of
virtual photon in the inclusive pion-induced Drell-Yan varies from
transverse to longitudinal as the longitudinal momentum fraction
$x_{\pi}$ of the parton inside the pion becomes
large~\cite{Anderson:1979xx,Conway:1989fs}. Specifically the dimuon
angular distribution changes from $(1+\cos^2 \theta)$ to
$\sin^2\theta$ as $x_{\pi} \to 1$. The annihilation of the on-shell
quark and antiquark pair implies the production of a transversely
polarized photon which leads to the $(1+\cos^2 \theta)$
distribution. The change in the polarization of the virtual photon could
be understood as the dominance of higher-twist contributions in the
forward production~\cite{Berger:1979du}, approaching the so called
``Berger-Brodsky'' limit, illustrated in Fig.~\ref{fig:edycross1},
where Feynman $x_{F}\equiv {q'}^-/{q}^- \to 1$ and ${q'}^2={Q'}^2 \to
\infty$ at fixed ${Q'}^2(1-x_{F})$ and
$\overrightarrow{q'_T}$~\cite{Hoyer:2008fp}. In this limit the invariant mass of
inclusive hadronic final state $M_X$ remains finite at ${Q'}^2 \to
\infty$ as follows:
\begin{equation}
M_X^2 = (q+p-q')^2 \simeq (1-x_B)[(1-x_F)s+ x_F m_N^2]-\overrightarrow{q'_T}^2
\end{equation}
where $s=(p+q)^2$ is the squared center-of-mass energy, $m_N$ the mass
of nucleon, $\overrightarrow{q'_T}$ the transverse component of ${q'}$ and
$x_{B}\equiv {q'}^+/{p}^+ \simeq {Q'}^2/(x_F s)$.

In the limit of large $x_{\pi}$, the annihilating $\bar{u}$ antiquark
from the pion, with the transverse momentum $k_T$, is highly
off shell, $p_{\bar u}^2=-{\vec
  k_T}^{\ 2}/(1-x_{\pi})$~\cite{Berger:1979du}, while the $u$ quark
from the nucleon is nearly on shell. The antiquark of the pion is
subject to the bound-state effects characterized by the distribution
amplitude and it could be resolved via a large-virtuality gluon
exchange with the spectator quark. The amplitude is expressed as a
convolution of the corresponding partonic amplitude with the pion
DA. Thus the Drell-Yan production induced by antiquarks with large
$x_{\pi}$ from pion can be viewed as a semi-exclusive one and the
angular distribution of the produced muon pair was shown to be
sensitive to the pion DAs~\cite{Brandenburg:1994wf,Bakulev:2007ej}.

When the highly virtual gluon exchanged between two valence quarks of
the pion has timelike momentum instead of spacelike momentum, the
leading-twist contribution may be obtained by neglecting the
transverse momentum $k_T$ of the gluon, such that the spectator quark
becomes collinear to the pion and may be absorbed by the remnant of
the target. When this ``complete annihilation'' of the pion
constituents is accompanied by the final hadronic state $X = N$, this
process turns into the ``exclusive'' Drell-Yan production $\pi N
\rightarrow \gamma^* N$ in the forward direction, as illustrated in
Fig.~\ref{fig:edycross2}. The exclusive Drell-Yan process with a small
momentum transfer to the nucleon could be realized in Berger-Brodsky
limit~\cite{Hoyer:2008fp}, and results in the twist-two mechanism to
make the relevant (annihilating) $\bar{u}$ quark in
Fig.~\ref{fig:edycross2} off shell, which leads to a
longitudinally polarized virtual photon associated with the
$\sin^2\theta$ angular distribution [see Eq.~(\ref{cadall}) below].

\subsection{Leading-order factorization formula}
\label{sec:exc_dy1}

As mentioned above, factorization has been proven for the DVMP
processes at the leading twist, including the exclusive
electroproduction of pion, $\gamma^* N \to \pi
N$~\cite{Collins:1996fb}. In the limit of the large photon virtuality
$Q^2 = -q^2 > 0$ at fixed scaling variable, the Bjorken $x_B = Q^2 /(2
p \cdot q)$ and invariant momentum transfer $t = (p-p')^2$, with $q$,
$p$ and $p'$ the momenta of the virtual photon, initial, and final
nucleons, respectively, the amplitude can be written in terms of the
hard-scattering processes at parton level, combined with the DA
$\phi_\pi$ describing the formation of the pion from a $q\bar{q}$
pair, and also the nucleon GPDs, $\tilde{H}$ and $\tilde{E}$.

Interchanging the initial and final states in the pion production, and
replacing the spacelike momentum of $\gamma^*$ by the timelike
momentum without affecting the factorization proof order by order in
perturbation theory, the factorization at twist-two is argued to be
applicable to the exclusive Drell-Yan process, $\pi(q) N(p) \to
\gamma^*(q') N(p')$, with the same universal nonperturbative
input~\cite{Berger:2001zn}. The appropriate kinematical region is of
large timelike virtuality $Q'^2 = q'^2$ at fixed $t=(p'-p)^2$ and
fixed scaling variable $\tau$, defined as
\begin{equation}
\tau = \frac{Q'^2}{2 p\cdot q} \approx \frac{Q'^2}{s-m_N^2} \approx
\frac{Q'^2}{s} ,
\label{Eq:tau}
\end{equation}
where $s=(p+q)^2$ is the squared center-of-mass energy and $m_N$ the
mass of nucleon. The variable $\tau$ plays a similar role as the
Bjorken variable $x_B$ in DVMP and DVCS induced by the spacelike
$\gamma^*$ and is related to the pion beam momentum $P_\pi \approx
s/(2m_N) \approx Q'^2/(2m_N\tau)$. The skewness variable $\xi$ defined
in Eq.~(\ref{eqn:xi}) becomes~\cite{Berger:2001xd}
\begin{equation}
\xi \approx \frac{Q'^2}{2s - Q'^2} = \frac{\tau}{2 - \tau},
\label{Eq:xi2}
\end{equation}

At the large $Q'$ scaling limit, the corresponding leading-twist cross
section of $\pi^- p \to \gamma^* n$ as a function of $t$ and $Q'^2$ is
expressed in terms of convolution integrals $\tilde{\cal H}^{du}$ and
$\tilde{\cal E}^{du}$, as follows~\cite{Berger:2001zn}
\begin{align}
\left.\frac{d\sigma_L}{dt dQ'^2}\right|_{\tau}
&= \frac{4\pi \alpha_{\rm em}^2}{27}\frac{\tau^2}{Q'^8} f_\pi^2\, \Bigl[ (1-\xi^2) |\tilde{\cal H}^{du}(\tilde{x},\xi,t)|^2 \nonumber \\
&- 2 \xi^2 \mbox{Re}\ \bigl( \tilde{\cal H}^{du}(\tilde{x},\xi,t)^* \tilde{\cal E}^{du}(\tilde{x},\xi,t) \bigr)
   -  \xi^2 \frac{t}{4 m_N^2}|\tilde{\cal E}^{du}(\tilde{x},\xi,t)|^2 \Bigr],
\label{eq_dcross}
\end{align}
where the scaling variable $\tilde{x}$ is given by~\cite{Diehl:2003ny,Belitsky:2005qn,GPD,Berger:2001xd}
\begin{equation}
\tilde{x}=-\frac{(q+q')^2}{2(p+p') \cdot (q+q')} \approx - \frac{Q'^2}{2s - Q'^2} = -\xi
\label{Eq:barx}
\end{equation}
and $f_\pi$ is the pion decay constant. The subscript ``$L$'' of the
cross section indicates the contribution due to the
longitudinally polarized virtual photon.

The convolution integral $\tilde{\cal H}^{du}$ involves two soft
objects: the GPD $\tilde{H}^{du}$ for $p\rightarrow n$ transition of
Eq.~(\ref{eqn:gpd-ppn}) and the twist-two pion DA $\phi_{\pi}$ of
Eq.~(\ref{eqn:qbar-q-matrixnew}). Using Eq.~(\ref{isospinrelation}) to
relate the transition GPD with the usual proton GPDs $\tilde{H}^q$ for
quark flavor $q=u,d$, the expression of $\tilde{\cal H}^{du}$ is
given, at the leading order in $\alpha_s$, by~\cite{Berger:2001zn}
\begin{align}
  \tilde{\cal H}^{du}(\tilde{x},\xi,t) &= \frac{8}{3} \alpha_s \int_{-1}^1
  dz\, \frac{\phi_\pi(z)}{1-z^2} \nonumber \\ 
  &\times \int_{-1}^1 dx
  \Bigl( \frac{e_d}{\tilde{x}-x- i\epsilon} - \frac{e_u}{\tilde{x}+x- i\epsilon}
  \Bigr) \bigl( \tilde{H}^{d}(x,\xi,t) - \tilde{H}^{u}(x,\xi,t)
  \bigr),
\label{eq_Hdu}
\end{align}
where $e_{u,d}$ are the electric charges of $u,d$ quarks in units of
the positron charge.  The corresponding expression of $\tilde{\cal
  E}^{du}$ is given by (\ref{eq_Hdu}) with $\tilde{H}^q$ replaced by
the proton GPDs $\tilde{E}^q$. Because of the pseudoscalar nature of the
pion, the cross section (\ref{eq_Hdu}) receives the contributions of
$\tilde{H}$ and $\tilde{E}$ only, among the GPDs in
Eqs.~(\ref{eqn:gpd-n}) and ~(\ref{eqn:gpd-p}). 

The leading-twist cross section (\ref{eq_dcross}) enters the four fold
differential cross sections for $\pi^- p \to \gamma^* n$
as~\cite{Goloskokov:2009ia,Goloskokov:2015zsa},
\begin{align}
\frac{d\sigma}{dt dQ'^2 d\cos\theta d\varphi} &= \frac{3}{8\pi} \bigl( \sin^2\theta \frac{d\sigma_L}{dt dQ'^2} + \frac{1+\cos^2\theta}{2} \frac{d\sigma_T}{dt dQ'^2} \nonumber \\ 
&+ \frac{\sin2\theta \cos\varphi}{\sqrt{2}} \frac{d\sigma_{LT}}{dt dQ'^2} + \sin^2\theta \cos2\varphi \frac{d\sigma_{TT}}{dt dQ'^2} \bigr),
\label{cadall}
\end{align}
with the angles $(\theta, \varphi)$ specifying the directions of the
decay leptons from $\gamma^*$. In Eq.~(\ref{cadall}) $d\sigma_T/ (dt
dQ'^2)$ is the cross section due to the transversely polarized virtual
photon while $d\sigma_{LT}/ (dt dQ'^2)$ and $d\sigma_{TT}/ (dt dQ'^2)$
are the longitudinal-transverse interference and transverse-transverse
(between helicity $+1$ and $-1$) interference contributions,
respectively. The $d\sigma_{LT}/ (dt dQ'^2)$ is of twist-three and is
suppressed asymptotically by $1/Q'$ compared to the twist-two cross
section $d\sigma_{L}/ (dt dQ'^2)$, while $d\sigma_T/ (dt dQ'^2)$ and
$d\sigma_{TT}/ (dt dQ'^2)$ are suppressed by one more power of $1/Q'$
as twist-four effects. The angular structures of these four terms,
characteristic of the associated virtual-photon polarizations, allow
us to separate the contribution of the leading-twist cross section
$d\sigma_{L}/ (dt dQ'^2)$ from the measured angular distributions.

\subsection{Pion-pole dominance and pion timelike form factor}

As shown in Eq.~(\ref{eq_dcross}), the term associated with
$|\tilde{\cal E}^{du}|^2 $ is multiplied by the momentum transfer
$|t|$ and thus the contribution of nucleon $\tilde{E}^{q}$ tends to be
suppressed at small $|t|$ compared to that of
$\tilde{H}^{q}$. Nevertheless, a remarkable feature of $\tilde{E}^{q}$
is that chiral symmetry ensures that $\tilde{E}^{q}(x,\xi,t)$ receives
a significant pion-pole contribution for the ERBL region $|x| \le
\xi$, and, therefore, $\tilde{E}^{q}$ could play an important role at
small $|t|$ due to the proximity of the pion pole at $|t|=m_\pi^2$.

In principle, $\tilde{E}^{q}$ has the pion-pole and non-pole
contributions, but it is demonstrated that the latter is small
compared with the former~\cite{Frankfurt:1999fp}. Thus,
$\tilde{E}^{q}$ is frequently parametrized in terms of the pion-pole
contribution as~\cite{Berger:2001xd,Kroll:2012sm}
\begin{equation}
\tilde{E}^{u}(x,\xi,t)=-\tilde{E}^{d}(x,\xi,t)=\Theta(\xi-|x|)\frac{F(t)}{2\xi}\phi_{\pi}(x/\xi),
\label{eqn:E_{tilde}}
\end{equation}
using the pion DA (\ref{eqn:qbar-q-matrixnew}), and $\Theta$ is
the step function. Here, the form factor $F(t)$ coincides with the
nucleon pseudoscalar form factor based on the moment sum rule
(\ref{sumrule}) applied to the pion-pole approximation to
$\tilde{E}^{q}$, so that its behavior near the pion pole is determined
by the partially conserved axial-vector current relation as $F(t\sim m_\pi^2) \simeq 4m_N^2g_A(0)/(m_\pi^2 -t)$, with
$g_A(t)$ the axial form factor of the nucleon. Using the
Goldberger-Treiman relation, $g_A(0)=f_\pi g_{\pi NN}/(\sqrt{2}m_N)$
($\approx 1.25$), with $g_{\pi NN}$ the pion-nucleon coupling
constant. Taking into account the form factor $\tilde{F}(t)$ for the
off-shell behavior, we may express $F(t)$ as
\begin{equation}
F(t) = m_N f_{\pi} \frac{2\sqrt{2}g_{\pi NN} \tilde{F}(t)}{m_{\pi}^2 - t} .
\label{psff}
\end{equation}
In Ref.~\cite{Goloskokov:2015zsa}, 
$\tilde{F}(t)$ is parametrized as the pion-nucleon vertex form factor,
\begin{equation}
\tilde{F}(t)=F_{\pi NN}(t) = \frac{\Lambda_N^2 -  m_{\pi}^2}{\Lambda_N^2 - t},
\label{ftilde1}
\end{equation}
where $\Lambda_N=0.44$~GeV~\cite{Goloskokov:2009ia}. A different form
according to the results in the chiral soliton model of the
nucleon~\cite{Frankfurt:1999fp} is used in Ref.~\cite{Berger:2001zn}:
\begin{equation}
\tilde{F}(t)=1-\frac{B(m_\pi^2 -t)}{(1-Ct)^2},
\label{ftilde2}
\end{equation}
with $B=1.7$~GeV$^{-2}$ and $C=0.5$~GeV$^{-2}$.

\begin{figure}[htbp]
\begin{center}
\centering
\includegraphics[width=0.5\textwidth]{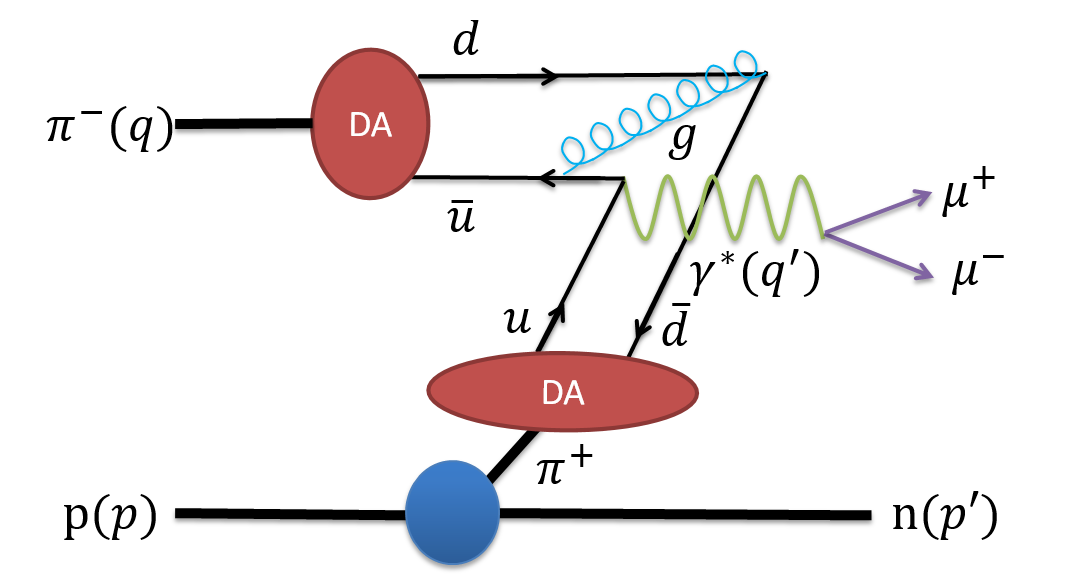}
\caption
[\protect{}]{Pion-pole contribution to the distribution $\tilde{E}$.}
\label{fig:pion_pole}
\end{center}
\end{figure}

As for the exclusive DY amplitude, the GPD $\tilde{E}^{q}$ with
  the expressions of Eqs.~(\ref{eqn:E_{tilde}}) and (\ref{psff}) gives
  rise to a contribution
\begin{equation}
\propto \frac{g_{\pi NN}\tilde{F}(t)}{m_\pi^2 -t}F^{\mbox{\scriptsize
    tw-2}}_{\pi}(Q'^2),
\label{ppc}
\end{equation}
which corresponds to the mechanism illustrated by
Fig.~\ref{fig:pion_pole}, where $F^{\mbox{\scriptsize
    tw-2}}_{\pi}(Q'^2)$ denotes the collinear factorization formula
for the timelike electromagnetic form factor of the pion, as the
convolution of the leading-order partonic hard-scattering amplitude, $\alpha_s
T^{(0)}_H(Q'^2)$, with the two DAs (\ref{eqn:qbar-q-matrixnew}) of
twist-2,
\begin{equation}
F^{\mbox{\scriptsize tw-2}}_{\pi}(Q'^2) \sim \phi_{\pi}
\otimes\alpha_s T^{(0)}_H(Q'^2) \otimes\phi_{\pi}.
\label{feltw2}
\end{equation}
Indeed, the contribution (\ref{ppc}) due to the GPD $\tilde{E}^{q}$
(\ref{eqn:E_{tilde}}) could produce a dominant effect at small
$|t|\sim 0$ due to the proximity of the pion pole, while the factor
$\tilde{F}(t)/(m_\pi^2-t)$ in Eqs.~(\ref{ftilde1}) or (\ref{ftilde2})
gives significant suppression of the contribution with an increase of
$|t|$.

Suppose we encounter an enhancement in the experimental data at small
$|t|\sim 0$ compared to the prediction based on Eq.~(\ref{ppc}); this
would be a sign of significant corrections beyond the twist-2 LO
factorization formula (\ref{feltw2}). Indeed, constraints from chiral
symmetry fix the form of the contributions (\ref{ppc}), such that the
possible higher-order contributions associated with the pion-pole
behavior at small $|t|\sim 0$ modify the pion electromagnetic timelike
form factor $F^{\mbox{\scriptsize tw-2}}_{\pi}(Q'^2)$ only. Taking
into account all such contributions, one would eventually obtain the
result (\ref{ppc}) with $F^{\mbox{\scriptsize tw-2}}_{\pi}(Q'^2)$
replaced by the full electromagnetic pion form factor
$F_\pi(Q'^2)$. In the formal limit $Q'^2 \rightarrow \infty$, we find
$F_\pi(Q'^2) \rightarrow F^{\mbox{\scriptsize tw-2}}_{\pi}(Q'^2)$,
asymptotically. But, $|F_\pi(Q'^2)| \gg |F^{\mbox{\scriptsize
    tw-2}}_{\pi}(Q'^2)|$ for $Q'^2 \lesssim$ a few GeV$^2$, i.e., the
intermediate values of our interest.

Considering the above complication known for the pion electromagnetic
form factor, we should apply the present factorization framework with
Eq.~(\ref{ppc}) to the measurements at large-$|t|$ regions, where the
pion-pole contribution is not dominating. Thus, information of the
nucleon GPDs and pion DAs could be extracted. At the same time, the
result using Eq.~(\ref{ppc}) with $F^{\mbox{\scriptsize
    tw-2}}_{\pi}(Q'^2)$ replaced by $F_\pi(Q'^2)$ may be compared with
the measurements restricted to the region $|t|\sim 0$; this approach
could offer a unique way to access the pion timelike form factor
$F_\pi (Q'^2)$, compared with the other measurement in, e.g., $e^+e^-
\rightarrow \pi^+\pi^-$.

It is noted that with an interchange of the initial and final states
in the above consideration, the GPD $\tilde{E}^{q}$ also produces a
dominant contribution in the deeply virtual pion production at
$|t|\sim 0$, giving rise to the contribution (\ref{ppc}) with
$F^{\mbox{\scriptsize tw-2}}_{\pi}(Q'^2)$ replaced by the
corresponding twist-2 factorization formula with spacelike $Q^2$. It
is found that the framework corresponding to the leading-twist
calculation of the pion form factor failed to describe the experimental
data of DVMP of $\pi^+$~\cite{Goloskokov:2009ia}, and the replacement
of $F^{\mbox{\scriptsize tw-2}}_{\pi}$ by an empirical behavior of the
pion electromagnetic form factor $F_\pi$ was favored by the data for
$|t|\sim 0$. Clarifying roles of the pion-pole contribution over a
wide range of $t$ requires calculations beyond the leading
mechanism~\cite{Braun:1999uj,Li:2010nn,Hu:2012cp}.

\subsection{Predicted differential cross sections of exclusive Drell-Yan process}
\label{sec:exc_dy2}

With the parametrization of $\tilde{H}^{q}(x,\xi,t)$ and
$\tilde{E}^{q}(x,\xi,t)$ GPDs and pion DAs, the LO differential cross
sections of the exclusive Drell-Yan process could be evaluated by
Eqs.~(\ref{eq_dcross}) and~(\ref{eq_Hdu}) straightforwardly. Since the
global analysis of GPDs is still at a premature stage, we use two sets
of GPD modeling to estimate the uncertainty due to the GPD input. In
terms of consistency, the same pion DA is used for the modeling of
$\tilde{E}^{u}-\tilde{E}^{d}$ and the convolution integrals of
$\tilde{\cal H}^{du}$ and $\tilde{\cal E}^{du}$.

The first set of GPDs labeled as ``BMP2001'' is what was used in
Refs.~\cite{Berger:2001zn,Berger:2001xd}, taking a factorizing ansatz
for the $t$ dependence, $\allowbreak \tilde{H}^{d,u}(x,\xi,t)
\allowbreak = \allowbreak \tilde{H}^{d,u}(x,\xi,0) \allowbreak
\left[g_A(t)/g_A(0)\right]$ and a dipole form assumed for the
$t$ dependence, $g_A(t)/g_A(0)$. Here, $\tilde{H}^q(x, \xi, 0)$ is
constructed from an ansatz based on double distributions as an
integral of $\tilde{H}^q(x, 0, 0)$ $=\Delta q(x)$ combined with a
certain profile function generating the skewness $\xi$
dependence~\cite{Berger:2001xd,Radyushkin:1998bz}. For $\Delta q(x)$,
a LO parametrization of polarized valence distributions is used. The
quantity $\tilde{E}^{u}-\tilde{E}^{d}$, arising in $\tilde{\cal
  E}^{du}$, is given as Eq.~(\ref{eqn:E_{tilde}}) with
Eqs.~(\ref{psff}) and (\ref{ftilde2}), in which the pion DA is taken
as the asymptotic form, $\phi_{\pi}(z)\rightarrow (3/4)(1-z^2)$,
corresponding to Eqs.~(\ref{eqn:qbar-q-matrixnew}) and
(\ref{normconnew}) with Eq.~(\ref{eqn:asymp}).

The second set of GPDs is constructed in a rather similar way as the
first one, and is labeled as ``GK2013''~\cite{Kroll:2012sm}. The
parameters are determined from the HERMES data on the cross sections
and target asymmetries for $\pi^+$
electroproduction~\cite{Airapetian:2007aa}. In addition the pion DA
used in the parametrization of $\tilde{E}^{u}-\tilde{E}^{d}$ as
Eq.~(\ref{eqn:E_{tilde}}) with Eqs.~(\ref{psff}) and (\ref{ftilde1})
is taken as $\phi_{\pi}(z)=(3/4)(1-z^2)[1+a_2 C_2^{(3/2)}(z)]$ with
$a_2~(\mu=2~{\rm GeV})=0.22$, i.e. the ``GK'' modeling in
Table~\ref{tab:pionDA}.

\begin{figure}[hbtp]
\begin{center}
\centering
\subfigure[]
{\includegraphics[width=0.8\textwidth]{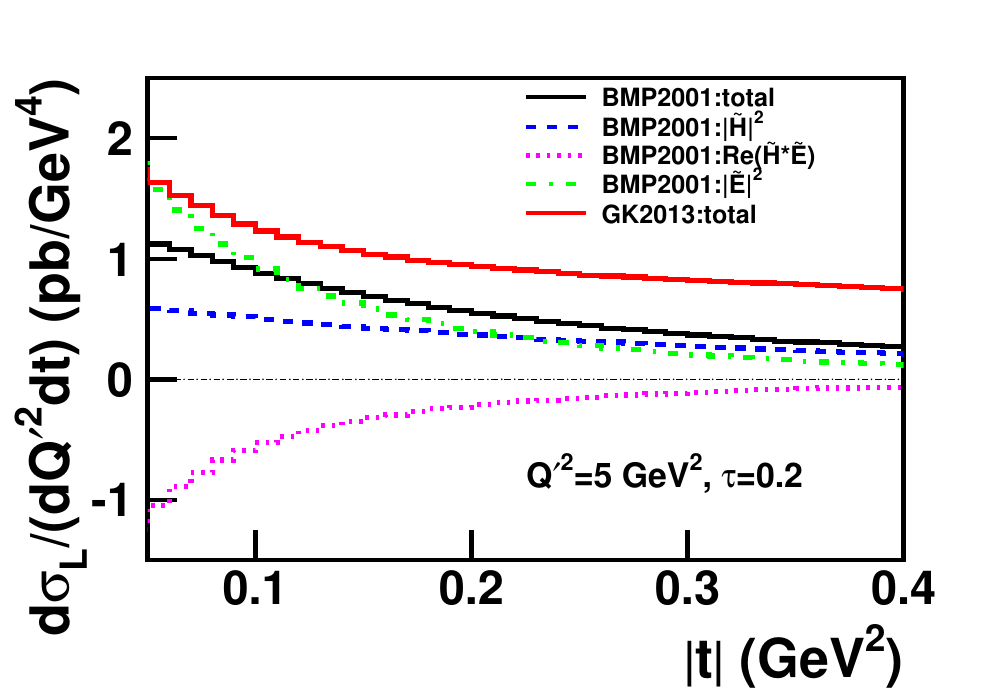}
\label{fig:edycrossPLB_1}}
\subfigure[]
{\includegraphics[width=0.8\textwidth]{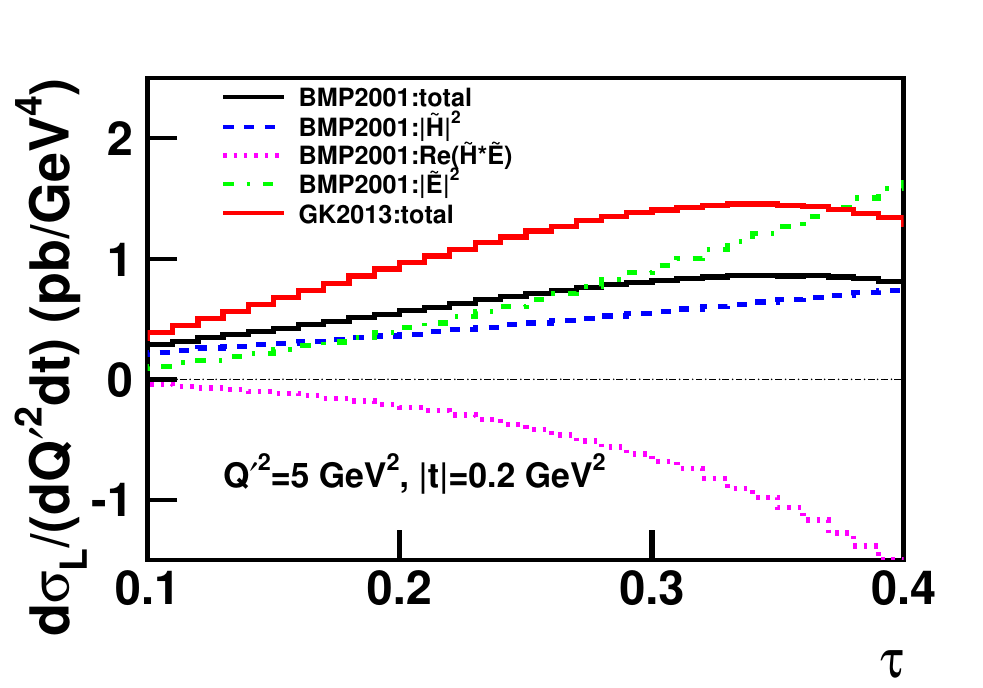}
\label{fig:edycrossPLB_2}}
\caption
[\protect{}] {(a) Differential cross section (\ref{eq_dcross}) of
  $\pi^- p \to \gamma^* n$ as a function of $|t|$ (full lines) for
  $Q'^2=5 \gev^2$ and $\tau = 0.2$. Individual contributions are shown
  for the terms with $|\tilde{\cal H}|^2$ (dashed), $\mbox{Re} (
  \tilde{\cal H}^*\, \tilde{\cal E} )$ (dotted), and $|\tilde{\cal
    E}|^2$ (dash-dotted). (b) Differential cross section
  (\ref{eq_dcross}) of $\pi^- p \to \gamma^* n$ as a function of
  $\tau$ (full lines) for $Q'^2=5 \gev^2$ and $|t|= 0.2
  \gev^2$.}
\label{fig:edycrossPLB}
\end{center}
\end{figure}

To check the correctness of our evaluation, we first repeat the
evaluation of the differential cross sections using the first set of
GPDs as what was done in Ref.~\cite{Berger:2001zn}. The differential
cross section (\ref{eq_dcross}) and its separate contributions from
the individual terms with $|\tilde{\cal H}|^2$, $\mbox{Re} (
\tilde{\cal H}^*\, \tilde{\cal E} )$, and $|\tilde{\cal E}|^2$ are
shown in Fig.~\ref{fig:edycrossPLB_1} as a function of the invariant
momentum transfer $|t|$ at $\tau=0.2$ and $Q'^2 =
5.0$~GeV$^2$. According to Eq.~(\ref{Eq:tau}), these kinematic
conditions correspond to $s = 26$~GeV$^2$, i.e. an interaction of
13~GeV pion beam with protons at rest. Fig.~\ref{fig:edycrossPLB_2} is
like Fig.~\ref{fig:edycrossPLB_1} but as a function of the scaling
variable $\tau$ at $|t|=0.2$~GeV$^2$.\footnote{A slight discrepancy
  between Fig.~\ref{fig:edycrossPLB_2} and a similar figure in
  Ref.~\cite{Berger:2001zn} is found, and we confirm that there is
  typo in the figure in Ref.~\cite{Berger:2001zn}.}  It is clear that
the relative importance of $\tilde{\cal H}$ and $\tilde{\cal E}$
depends on $|t|$ and $\tau$.

The results of $|t|$ and $\tau$ dependence of differential cross
sections under the same kinematic conditions with GK2013 GPDs are
shown as the red solid lines in Fig.~\ref{fig:edycrossPLB}. The
corresponding cross sections are slightly greater than those with
BMP2001 GPDs. Both production cross sections are of the order of a
few pb in the forward direction and this suggests very challenging
measurement of the exclusive Drell-Yan process.

\begin{figure}[htbp]
\begin{center}
\centering
\subfigure[]
{\includegraphics[width=0.48\textwidth]{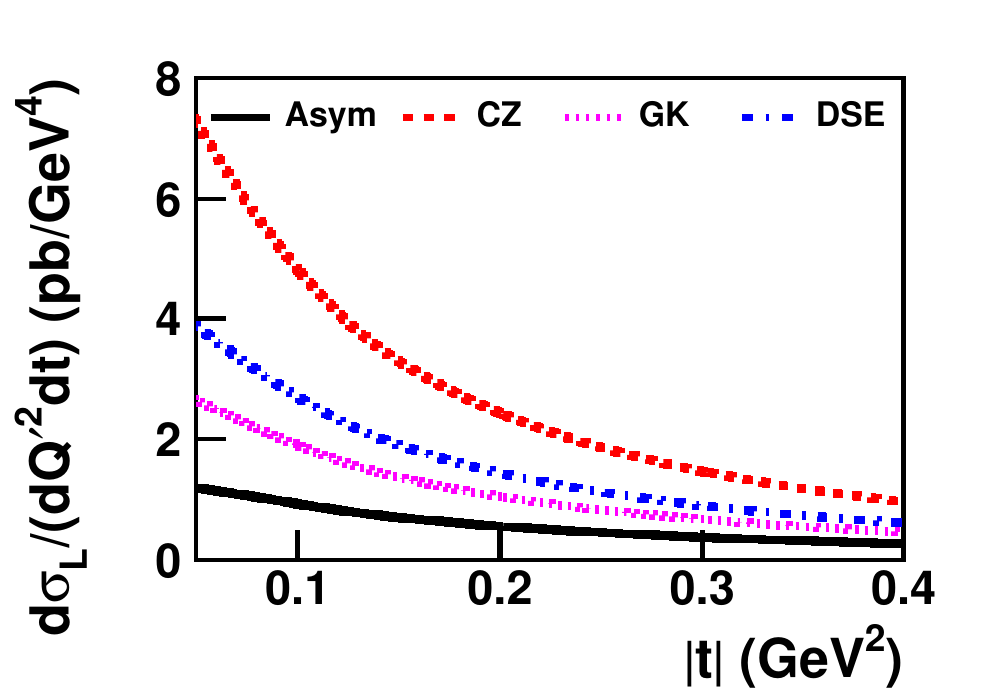}
\label{fig:edycross32}}
\subfigure[]
{\includegraphics[width=0.48\textwidth]{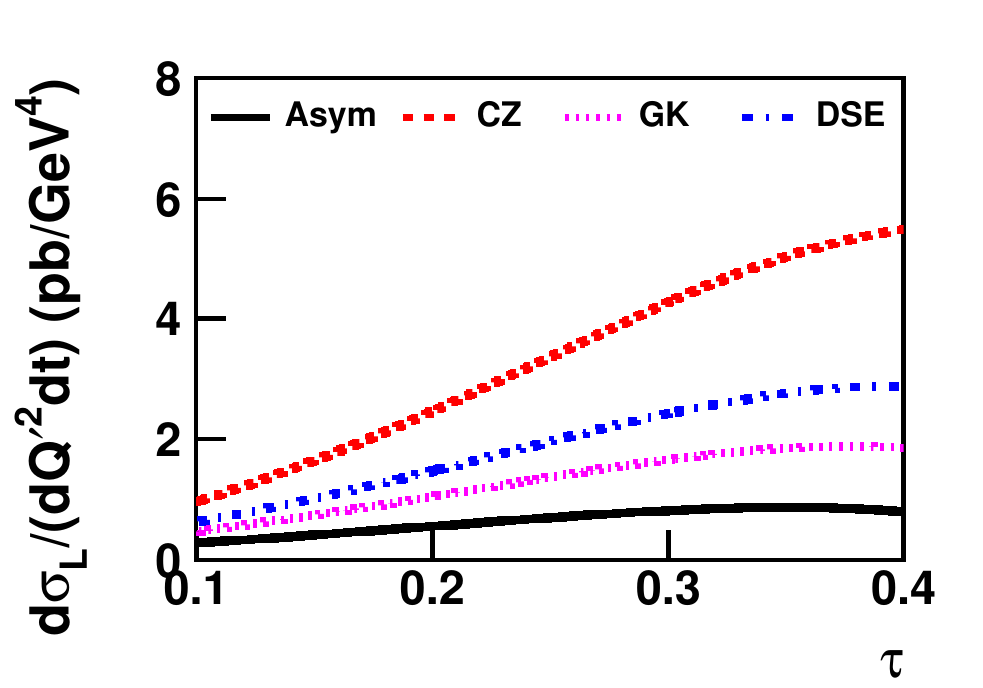}
\label{fig:edycross33}}
\caption
[\protect{}] {(a) Differential cross section (\ref{eq_dcross}) of
  $\pi^- p \to \gamma^* n$ as a function of $|t|$ at $Q'^2=5 \gev^2$
  and $\tau = 0.2$ for different pion DAs. (b) Differential cross
  section (\ref{eq_dcross}) of $\pi^- p \to \gamma^* n$ as a function
  of $\tau$ at $Q'^2=5 \gev^2$ and $|t|= 0.2 \gev^2$ for different
  pion DAs.}
\label{fig:pida}
\end{center}
\end{figure}

The QCD evolution effect on the GPDs~\cite{Vinnikov:2006xw} and pion
DAs is found rather minor in the present kinematic region. With the
BMP2001 GPDs we replace the pion DAs $\phi_\pi$ in the integrals
of $\tilde{\cal H}^{du}$ and $\tilde{\cal E}^{du}$ (\ref{eq_Hdu}) and
the parametrization of $\tilde{E}^{u}-\tilde{E}^{d}$
(\ref{eqn:E_{tilde}}) with different modeling in
Table~\ref{tab:pionDA}~\cite{erbl,pionDA:CZ,Kroll:2012sm,pionDA:DSE}
and the obtained production cross sections of the exclusive Drell-Yan
process show a strong sensitivity to the input of pion DAs as shown in
Fig.~\ref{fig:pida}. This suggests that such measurement will provide
another important way of constraining pion DAs other than the
determination of the pion-photon transition form factor in $e^+e^-$
collisions~\cite{Aubert:2009mc,Uehara:2012ag}.

Recently Goloskokov and Kroll extended the leading-order factorization
formula of the exclusive Drell-Yan process using the so-called
``modified perturbative approach''~\cite{Goloskokov:2015zsa},
retaining some effects of quark transverse momenta inside the pion,
and also relying on a different treatment of the pion-pole term which
gives the dominant contribution to $\tilde{E}^{u}-\tilde{E}^{d}$
arising in $\tilde{\cal E}^{du}$ of (\ref{eq_dcross}). They treated
the pion pole term, separately from the factorization framework, as
the hadronic one-particle-exchange amplitude combined with the
experimental values $F_\pi(Q'^2)$ of the timelike pion form factor,
making the replacement $F^{\mbox{\scriptsize
    tw-2}}_{\pi}(Q'^2)\rightarrow F_\pi(Q'^2)$ in the contribution
(\ref{ppc}). It was found that the forward production cross section
was enhanced by about a factor of 40, compared to the calculation of
(\ref{eq_dcross}) obtained with the input
BMP2001~\cite{Berger:2001zn}. The dominant contribution to the
large enhancement factor is from $|\tilde{\cal E}|^2$. This is mainly
due to the use of the experimental values of the pion form factor with
$Q'^2 |F_\pi(Q'^2)|\simeq 0.88$~GeV$^2$ instead of the leading-twist
result ($Q'^2|F^{\mbox{\scriptsize tw-2}}_{\pi}(Q'^2)|\simeq
0.15$~GeV$^2$). In this framework, Goloskokov and Kroll also gave an
estimate of the cross sections for the production of transverse
virtual photon in the exclusive Drell-Yan process, which leads to the
angular distributions shown in Eq.~(\ref{cadall}), demonstrating that
those cross sections are parametrized by the quark-helicity-flipping
GPDs $H_T, \tilde{H}_T, E_T$, and
$\tilde{E}_T$~\cite{Goloskokov:2009ia,Ahmad:2008hp}, and the higher-twist pion DAs.

%%%%%%%%%%%%%%%%%%%%%%%%%%%%%%%%%%%%%%
\section{Feasibility study of measurement at J-PARC}
\label{sec:feasibility}
%%%%%%%%%%%%%%%%%%%%%%%%%%%%%%%%%%%%%%

\subsection{High-momentum beam line and E50 experiment in J-PARC Hadron Hall}
\label{sec:JPARC}

J-PARC consists of three accelerator stages, the proton linear
accelerator of 400 MeV, the rapid-cycle 3 GeV proton synchrotron, and
the main 50-GeV proton synchrotron. An important feature of the J-PARC
accelerator is the high intensity of the primary proton
beam~\cite{J-PARC}. The interaction of the 30-GeV primary proton beam
with a production target provides high-intensity secondary beams of
pions, kaons, and antiprotons. There are three (plus one branched)
beam lines located in the Hadron Hall where nuclear and particle
physics experiments are carried out for the study of hypernuclei,
exotic hadrons and rare kaon decay~\cite{HadronHall}.

A high-momentum beam line is under construction at the Hadron
Hall. Branching from the main proton beam of $10^{13}$ or
$10^{14}$/sec, this beam line can transport the primary proton beam
with an intensity of $10^{10}$-$10^{12}$/sec to the Hadron Hall. In
addition, by installing a thin production target at the branching
point, one can obtain unseparated secondary beams such as pions,
kaons, and antiprotons. A beam swinger is planned for extracting
secondary beam so that high-flux zero-degree extraction can be
achieved for negative-charged hadrons.

At fixed extraction angle, the beam intensity of these secondaries
depends on the species and momentum. The momentum profile of secondary
beams and a typical intensity for 10$-$20~GeV pions would be in the
order of $10^{7}-10^{8}$/sec. The intensity for higher momentum
beams would be lower. A beam momentum resolution of better than 0.1\%
can be obtained by using the dispersive method. The beam line
construction is expected to be finished in 2018.

The E50 experiment at J-PARC plans to investigate charmed-baryon
spectroscopy via the measurement of a $\pi^- + p \rightarrow Y_{c}^{*} +
D^{*-}$ reaction at the high-momentum beam
line~\cite{E50_JPARC}. Spectroscopy of $Y_{c}^{*}$ could reveal the
essential role of diquark correlation in describing the internal
structure of hadrons. The mass spectrum of $Y_{c}^{*}$ will be
constructed by the missing-mass technique following the detection of
$D^{*-}$ via its charged decay mode. Large acceptance for charged
hadrons together with good momentum resolution are required. The E50
experiment received the stage-1 approval in 2014.

\begin{figure}[hbtp]
\centering
\includegraphics[width=\textwidth]{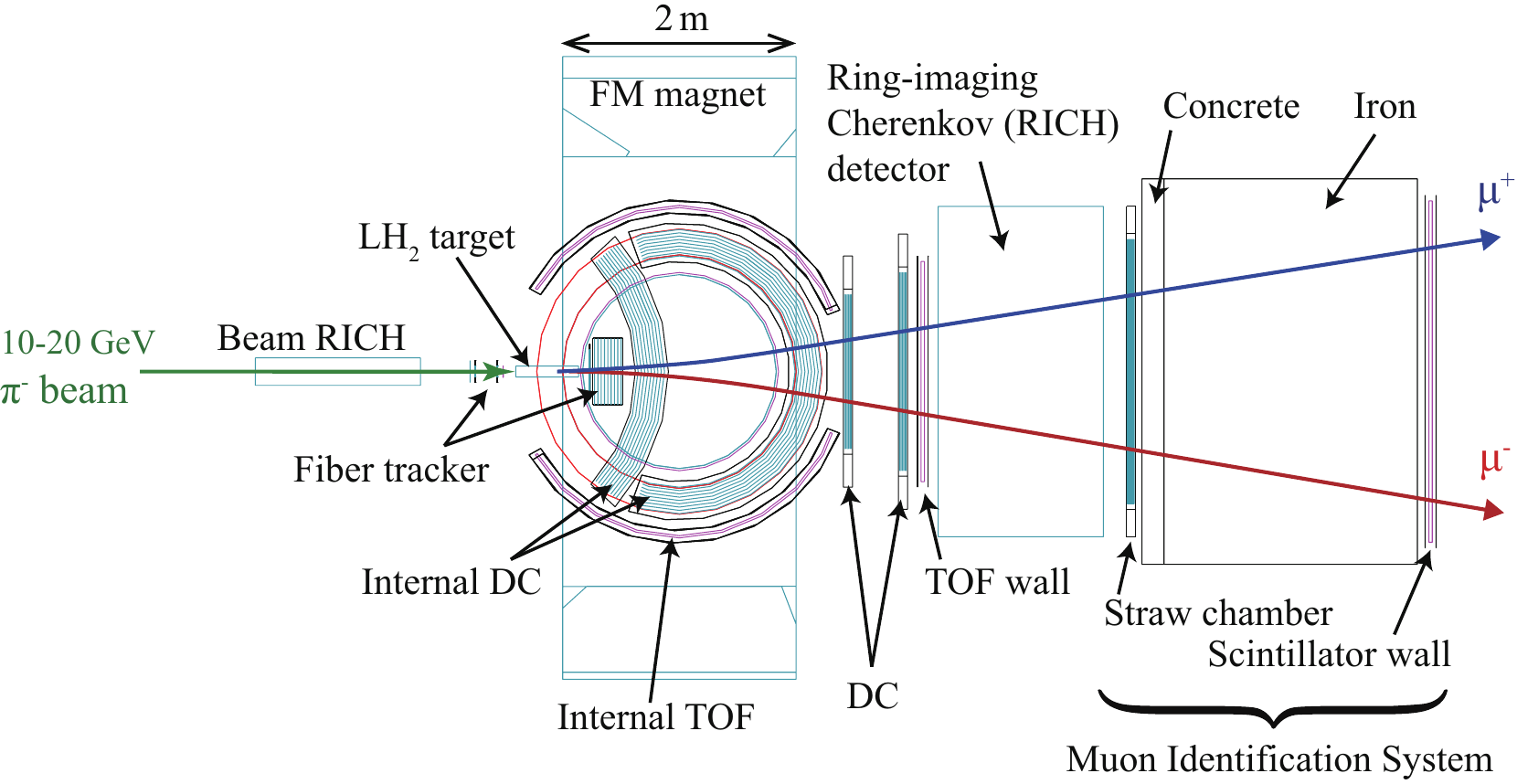}
\caption{Conceptual design of J-PARC E50 spectrometer with muon
  identification system.}
\label{fig:E50}
\end{figure}

Figure~\ref{fig:E50} shows the conceptual design of the E50
spectrometer. The spectrometer is composed of a dipole magnet and
various particle detectors~\cite{E50_JPARC}. Since the secondary beams
are unseparated, beam pions are tagged by gas Cherenkov counters (Beam
RICH) placed upstream of the target. Finely segmented particle
trackers, silicon strip detectors and scintillating fiber trackers
with designed spatial resolutions of 80 $\mu$m and 1 mm, respectively,
are placed immediately upstream and downstream of the target. The
magnet has a circular pole of 2.12 m in diameter and a gap of 1 m. An
integrated magnetic field of up to 2.3 Tesla-meter is expected.

High-granularity drift chambers placed downstream of the magnet are
for detection of charged tracks, e.g. kaons and pions from $D^{*-}$
decay. TOF counters and ring-imaging Cherenkov counters are placed
downstream of the drift chambers for high-momentum kaon/pion
separation. In the current spectrometer configuration, a missing-mass
resolution of $D^{*-}$ is expected to be as good as 5
MeV~\cite{E50_JPARC}.

Conventionally, the measurement of the Drell-Yan process in the
fixed-target experiments requires a hadron absorber immediately after
the targets to avoid large track densities in the spectrometer. Thanks
to the relatively low track density at the energy regime of J-PARC and
high-granularity tracking chambers, the measurement of the Drell-Yan
process could be operated without the installation of the hadron absorber
in front of the spectrometer. Exclusion of the multiple-scattering
effect in the hadron absorber is very essential for achieving a good
momentum determination of muon tracks so that the exclusive Drell-Yan
process can be characterized via the missing-mass technique. As for
the final muon identification, we propose to install a dedicated muon
identification ($\mu$ID) system in the most downstream position as
shown in Fig.~\ref{fig:E50}.

\subsection{Feasibility study}
\label{sec:measurement}

The features of large-acceptance and superb momentum resolution for
E50 spectrometer are suitable and essential for measuring the
exclusive pion-induced Drell-Yan process with missing-mass
technique. We perform the feasibility study using this detector
configuration together with $\mu$ID system in the Geant4 simulation
framework~\cite{Agostinelli:2002hh}. Both inclusive and exclusive
Drell-Yan events are generated together with the other dimuon sources
like $J/\psi$ and the random combinatorial from minimum-bias hadronic
events in the event simulation. We investigate if a signature of
exclusive Drell-Yan events could be clearly identified in the
missing-mass spectrum of reconstructed dimuon events. Below we
describe the details of the simulation and present the results of this
feasibility study.

\subsubsection{Kinematics}

The physics variables for characterizing the exclusive pion-induced
Drell-Yan process include the timelike virtuality $Q'$,
momentum-transfer squared $t$ and scaling variable
$\tau$. Experimentally, a pion beam of momentum $P_{\pi}$ collides with
a proton target and the momenta of produced $\mu^+$ and $\mu^-$ are
measured in the spectrometer for investigating this process. The
relationships between the kinematic variables $Q'$, $t$, and $\tau$,
and the experimental quantities $P_{\pi}$, $\vec{P}_{\mu^+}$, and
$\vec{P}_{\mu^-}$ are briefly reviewed.

First the timelike virtuality $Q'$ of virtual photon is simply the
invariant mass of dimuon $M_{\mu^{+}\mu^{-}}$ as illustrated in
Fig.~\ref{fig:edycross2}. The center-of-mass energy squared $s$ of the
two-body collision is fixed by the momentum $P_{\pi}$ (or the energy
$E_{\pi}$) of the pion beam as $s = m_{\pi}^2 + m_{N}^2 + 2 E_{\pi}m_{N}
\approx 2 P_{\pi}m_{N}$. From Eq.~(\ref{Eq:tau}) $\tau$ is related to
$P_{\pi}$ and $M_{\mu^{+}\mu^{-}}$ as $\tau = M_{\mu^{+}\mu^{-}}/(2
P_{\pi}m_{N})$.

The momentum transfer squared $t$ is related to the scattering angle
in the center-of-mass system $\theta^{CM}$ which can be determined by
the boost-invariant transverse momentum of the dimuon $P_{T \gamma^*}$
as follows:
\begin{align}
\label{Eq:kine_t}
t & = t_{0}-4 P_{\pi}^{CM} P_{\gamma^*}^{CM} \sin^2(\frac{\theta^{CM}}{2}) \\
\label{Eq:kine_pt}
P_{T \gamma^*} & = P_{T \gamma^*}^{CM} ={P_{\gamma^*}^{CM}} \sin(\theta^{CM}) \\
\label{Eq:kine_epi}
E_{\pi}^{CM} & = \frac{s+m_{\pi}^2-m_{N}^2}{2\sqrt{s}} \\
\label{Eq:kine_eg}
E_{\gamma^*}^{CM} & = \frac{s+M_{\mu^{+}\mu^{-}}^2-m_{N}^2}{2\sqrt{s}}
\end{align}
where $t_{0}$ ($= -4 m_N^2 \xi^2/(1-\xi^2)$) is the limiting value of
$t$ at $\theta^{CM}=0$.

\subsubsection{Exclusive Drell-Yan and background events}
\label{sec:MCevent}

\begin{figure}[hbtp]
\begin{center}
\centering
\subfigure[]
{\includegraphics[width=0.48\textwidth]{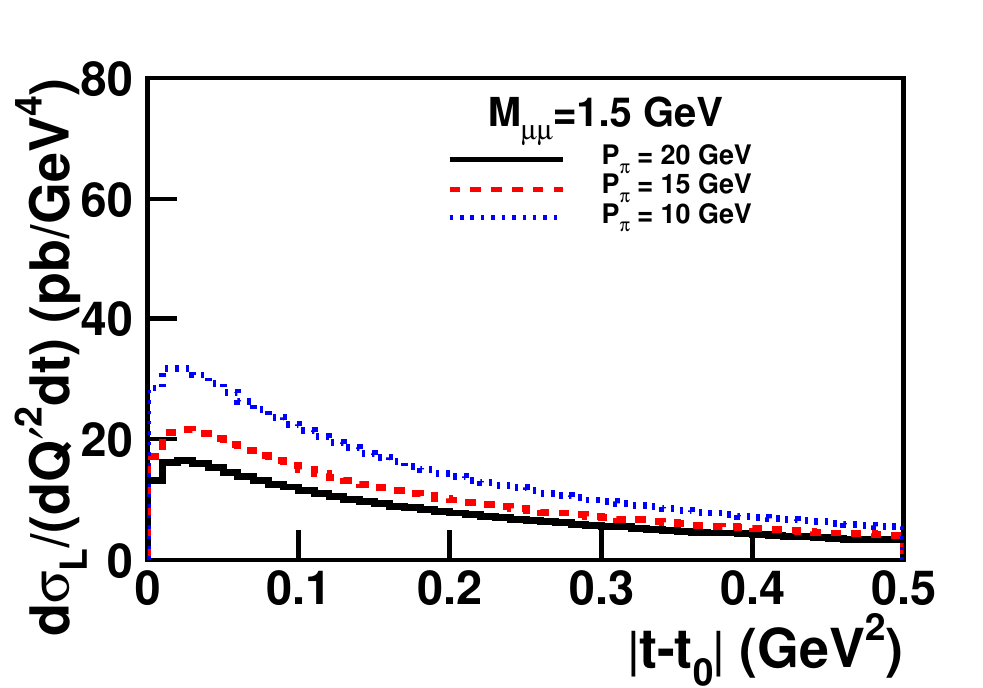}
\label{fig:edy_dcross1_1}}
\subfigure[]
{\includegraphics[width=0.48\textwidth]{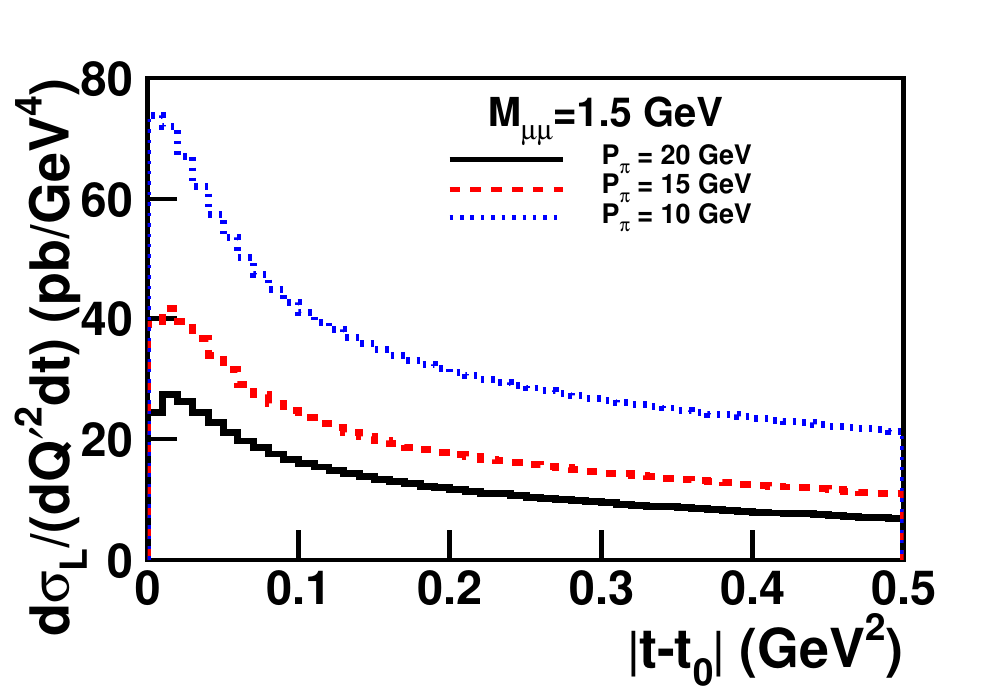}
\label{fig:edy_dcross1_3}}
\caption
[\protect{}]{Differential cross sections of exclusive Drell-Yan
  events, Eq.~(\ref{eq_dcross}), as a function of $|t-t_{0}|$ at
  $M_{\mu^{+}\mu^{-}}$=1.5 GeV for $P_{\pi}$ =10, 15 and 20 GeV with
  the input GPDs: (a) BMP2001 and (b) GK2013.}
\label{fig:edy_dcross1}
\end{center}
\end{figure}

\begin{figure}[hbtp]
\begin{center}
\centering
\subfigure[]
{\includegraphics[width=0.48\textwidth]{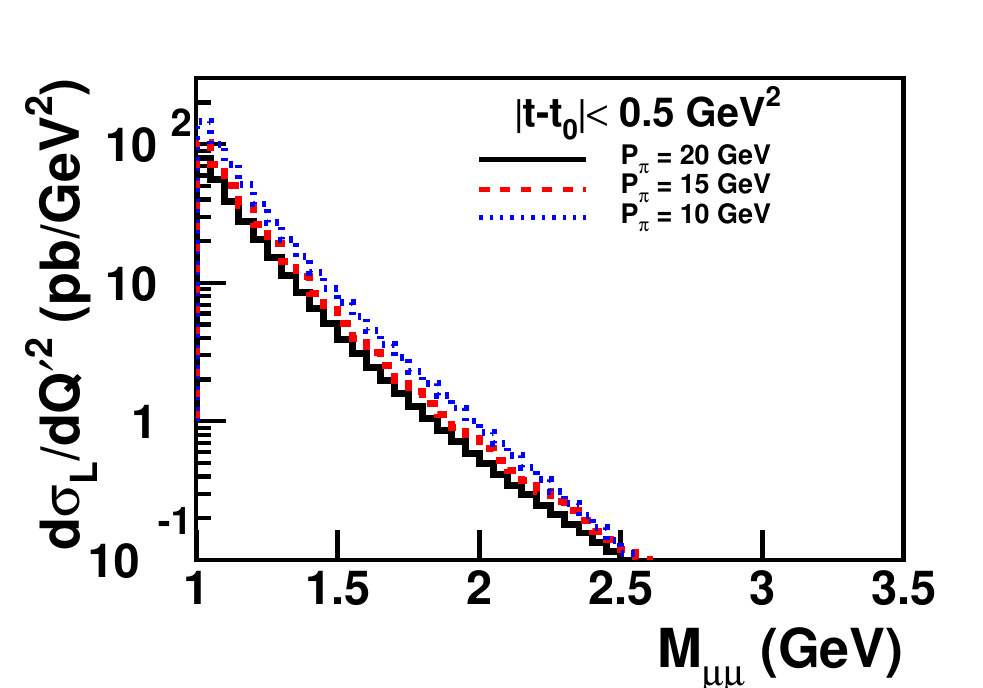}
\label{fig:edy_dcross2_1}}
\subfigure[]
{\includegraphics[width=0.48\textwidth]{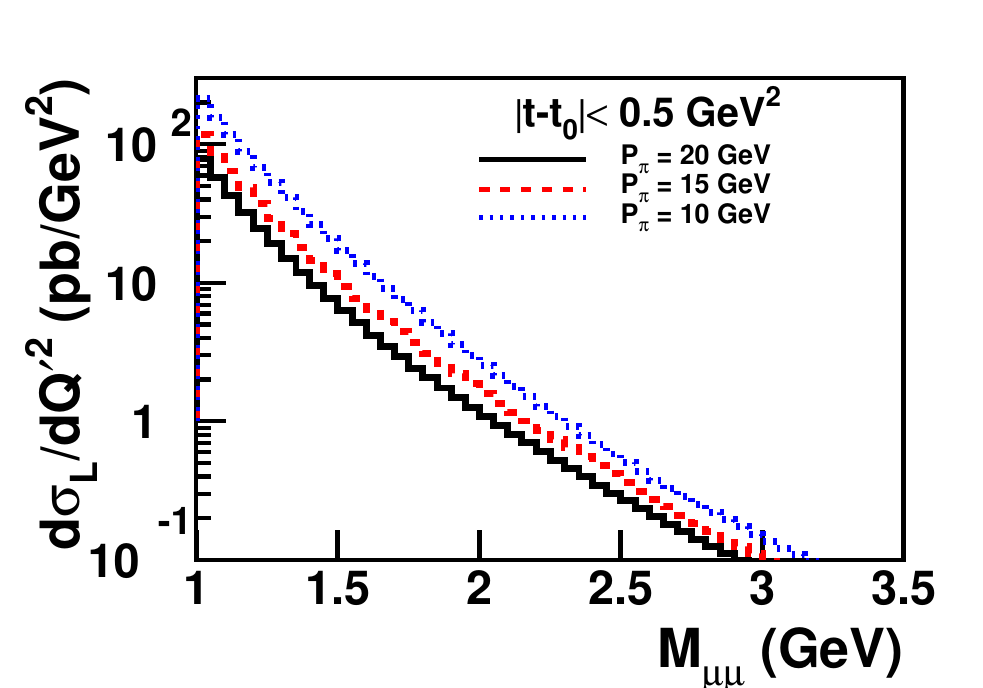}
\label{fig:edy_dcross2_3}}
\caption
[\protect{}] {Differential cross sections of exclusive Drell-Yan
  events, Eq.~(\ref{eq_dcross}), as a function of $M_{\mu^{+}\mu^{-}}$
  with $|t-t_0|<0.5$ GeV$^2$ for $P_{\pi}$ =10, 15 and 20 \gev~with
  the input GPDs: (a) BMP2001 and (b) GK2013.}
\label{fig:edy_dcross2}
\end{center}
\end{figure}

With the above two sets of GPDs, we obtain the differential cross
sections of exclusive Drell-Yan events, Eq.~(\ref{eq_dcross}), as a
function of $|t-t_0|$ at $M_{\mu^{+}\mu^{-}}=1.5$ GeV and those as a
function of $M_{\mu^{+}\mu^{-}}$ with $|t-t_0|<0.5$ GeV$^2$ in the
range of pion beam momentum $P_{\pi}$ = 10$-$20 GeV. The results for
$P_{\pi}$ = 10, 15 and 20 \gev~ are shown in
Figs.~\ref{fig:edy_dcross1} and ~\ref{fig:edy_dcross2}. As expected,
the production cross sections increase with smaller $|t|$,
$M_{\mu^{+}\mu^{-}}$, or $P_{\pi}$. The cross sections with GK2013
GPDs is about a factor of 2 larger than those with BMP2001 ones.

\begin{figure}[hbtp]
\begin{center}
\centering
\subfigure[]
{\includegraphics[width=0.48\textwidth]{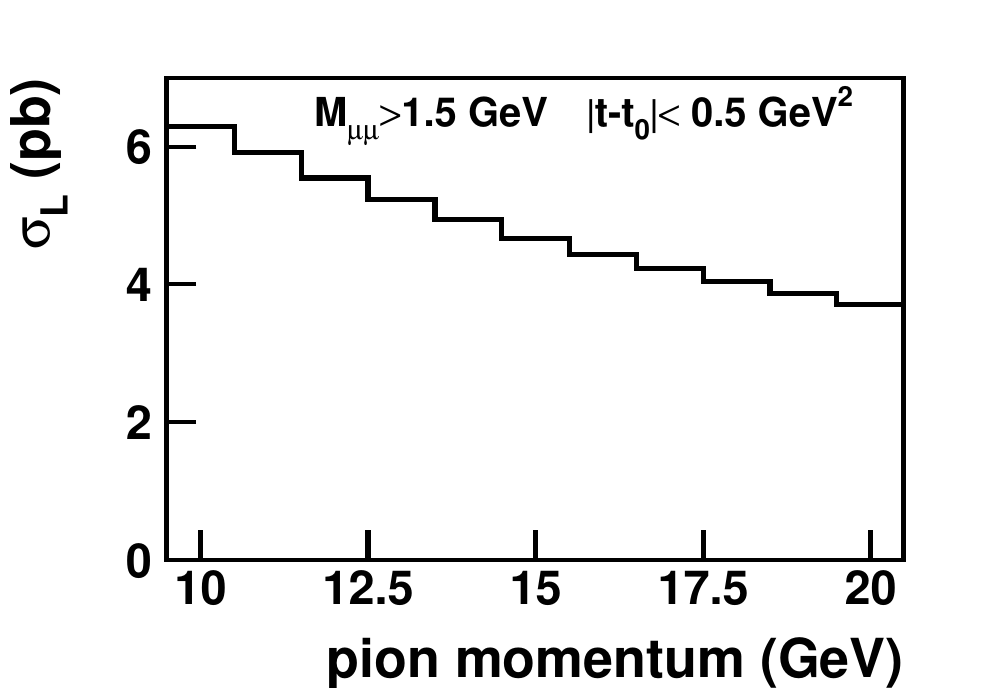}
\label{fig:edy_dcross3_1}}
\subfigure[]
{\includegraphics[width=0.48\textwidth]{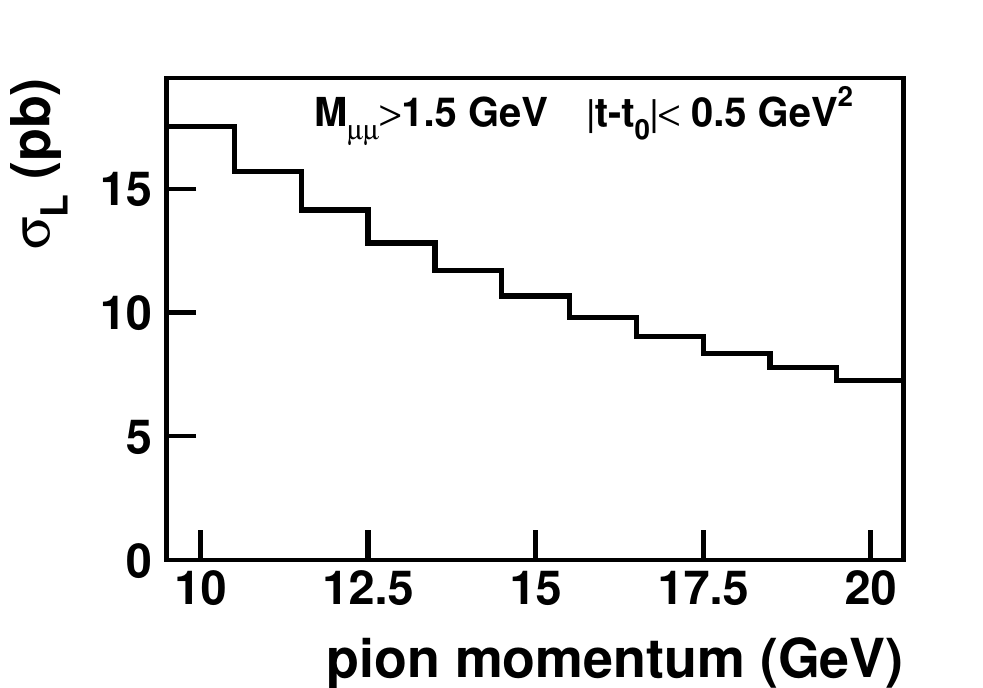}
\label{fig:edy_dcross3_3}}
\caption
[\protect{}] {Total leading-order cross sections of exclusive
  Drell-Yan events as a function of pion momentum $P_{\pi}$ in
  $M_{\mu^{+}\mu^{-}} > 1.5$ GeV and $|t-t_{0}| < 0.5$ GeV$^2$ with
  the input GPDs: (a) BMP2001 and (b) GK2013.}
\label{fig:edy_dcross3}
\end{center}
\end{figure}

For the feasibility study, we limit the exclusive/inclusive Drell-Yan
events in the invariant mass region $M_{\mu^{+}\mu^{-}} > 1.5$ GeV to
avoid the large combinatorial background in the low-mass region. The total
cross sections as a function of pion beam momentum are shown in
Fig.~\ref{fig:edy_dcross3}. As expected the cross sections of
exclusive hard processes drop as beam momentum increases. The total
cross section for $M_{\mu^{+}\mu^{-}} > 1.5$ GeV and $|t-t_{0}| < 0.5$
GeV$^2$ at J-PARC energies is about 5$-$10 pb in the current estimation.

Two major sources of background events have been considered: inclusive
Drell-Yan events and combinatorial background due to accidental
coincidence of dimuon in the final state. The inclusive pion-induced
Drell-Yan events are generated by PYTHIA 6 event
generator~\cite{Sjostrand:2006za} with the input of ``GRVPI1'' pion
PDF and ``CTEQ66'' proton PDF. The QCD K-factor is evaluated by
DYNNLO~\cite{DYNNLO} package. The production cross sections as a
function of invariant mass $M_{\mu^+ \mu^-}$ and dimuon transverse
momentum $p_T$ from the 16~GeV pion-induced Drell-Yan
data~\cite{Alspector:1978vb,McCal:1979zm} could be nicely
described. The hadronic background events with at least one single
muon in the final state are obtained from the low-energy JAM event
generator~\cite{JAM} and accidental background was constructed from
them. In addition we include $J/\psi$ production as background events.

The estimated total cross sections for the exclusive and inclusive
Drell-Yan events for the dimuon mass $M_{\mu^{+}\mu^{-}} > 1.5$ GeV
and the $|t-t_{0}|<0.5$ GeV$^2$ are summarized in
Table~\ref{tab:cross_table}. In the range of beam momentum 10$-$20 GeV,
the total hadronic interaction cross sections of $\pi^- p$ is about
20$-$30 mb while the production of $J/\psi$ is about 1$-$3 nb.

\begin{table}[hbt]
\begin{center}
\caption{ Expected cross sections for the exclusive and inclusive
  Drell-Yan processes.}
\scalebox{1.0}{
  \begin{tabular}{|c||c|c||c|}
  \hline
  & \multicolumn{2}{|c|}{Exclusive Drell-Yan} & Inclusive Drell-Yan \\
  & \multicolumn{2}{|c|}{
$\left(
\begin{array}{l}
M_{\mu^{+}\mu^{-}} > 1.5  $ GeV$, \\
|t-t_{0}|<0.5  $ GeV$^2
\end{array}
\right)$
} & ($M_{\mu^{+}\mu^{-}} > 1.5 $ GeV) \\
\cline{2-3}
  & {BMP2001} & {GK2013}  &     \\
  \hline
    $P_{\pi}$= 10 GeV  &   6.29 pb  &  17.53 pb   &   2.11 nb  \\
  \hline
    $P_{\pi}$= 15 GeV  &   4.67 pb  &  10.65 pb   &   2.71 nb  \\
  \hline
    $P_{\pi}$= 20 GeV  &   3.70 pb  &   7.25 pb   &   3.08 nb  \\
  \hline
  \end{tabular}
}
\label{tab:cross_table}
\end{center}
\end{table}

\subsubsection{Identification of the exclusive Drell-Yan events in missing-mass spectrum}
  
The $\mu$ID system is designed to consist of hadron absorber layers
made of 20-cm concrete and 230-cm iron to absorb incoming hadrons,
scintillator hodoscopes downstream of the absorber, and a straw
chamber (or drift tube chamber) upstream of the absorber. The
thickness of concrete and iron is optimized with the consideration of
the stopping power for low-momentum tracks and penetrating efficiency
for high-momentum ones. Muons are identified when the tracks can be
reconstructed in both sets of stations. The threshold momentum of a
penetrating muon is 3 GeV. The signals from $\mu$ID system could be
used in the trigger decision. Overall the $\mu$ID system is a minor
extension of the E50 spectrometer.

The major component of the background events is originated from
uncorrelated muons from the decay of hadrons, mostly pions and
kaons. These background muons arising from the decay of hadrons
produced on the target, could be effectively identified by a kink of
the decay vertex, bad $\chi^{2}$-probability in the reconstruction,
and inconsistency of the trajectory between the spectrometer and the
upstream tracking chamber in $\mu$ID system. The muons from the beam
decay could be rejected with proper kinematics cuts.
  
We perform the Monte-Carlo simulation assuming the following
experimental conditions: 4 g/cm$^2$ liquid hydrogen target,
$1.83/1.58/1.00 \times 10^7$ $\pi^{-}$/spill for $10/15/20$ GeV beam,
and a 50-day beam time. The expected rate of dimuon trigger from
the $\mu$ID system is $\sim$ $2/10/15$ Hz for $10/15/20$ GeV beams. The
corresponding integrated luminosity is $3.66/3.16/2$ fb$^{-1}$. Since
the trigger rate is low, the measurement of exclusive Drell-Yan
process and charmed-baryon spectroscopy could be carried out together
in the E50 experiment.

\begin{figure}[hbtp]
\centering
\hspace{-0.5cm}
\subfigure[]
{\includegraphics[width=0.33\textwidth]{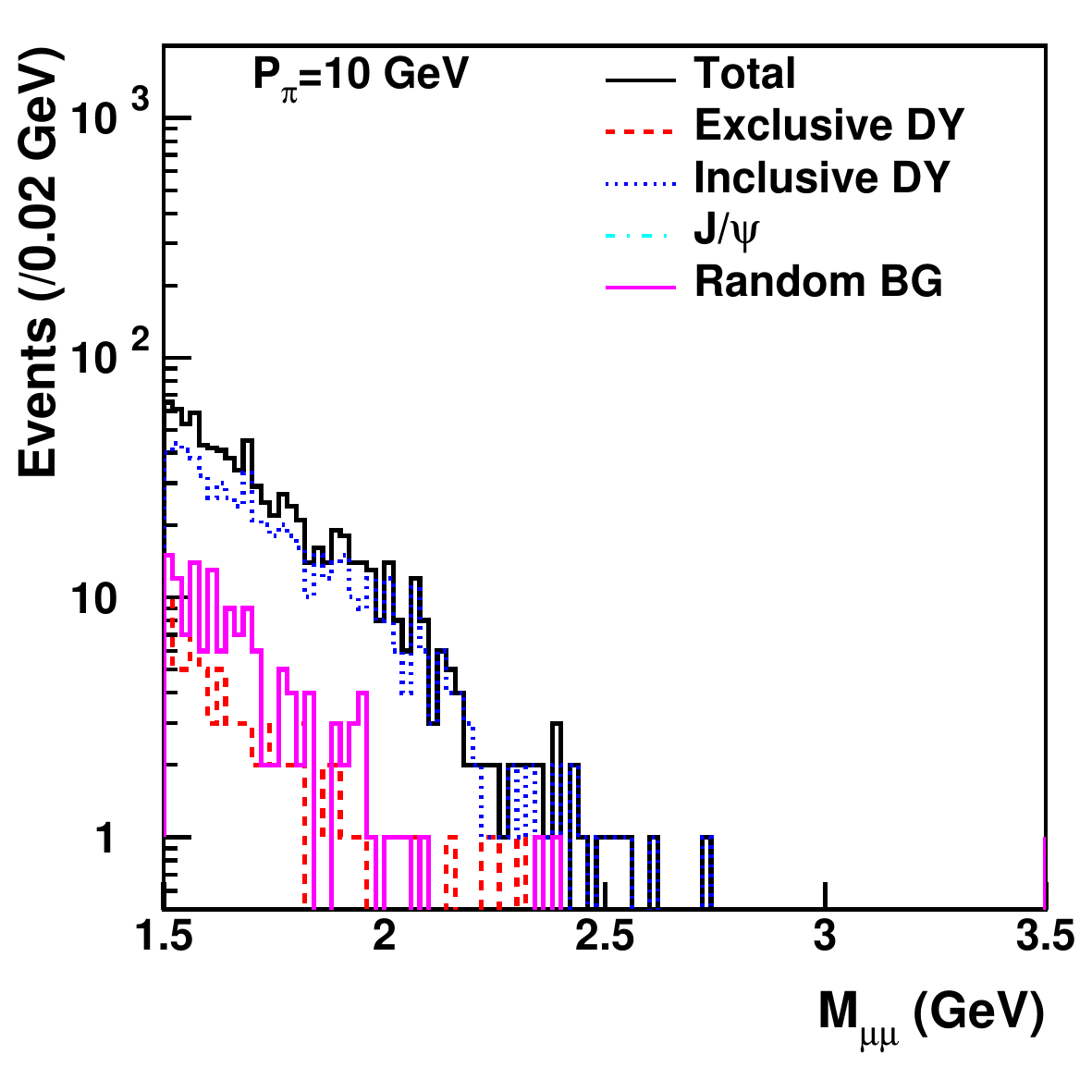}
\label{fig:invm_1}}
\hspace{-0.5cm}
\subfigure[]
{\includegraphics[width=0.33\textwidth]{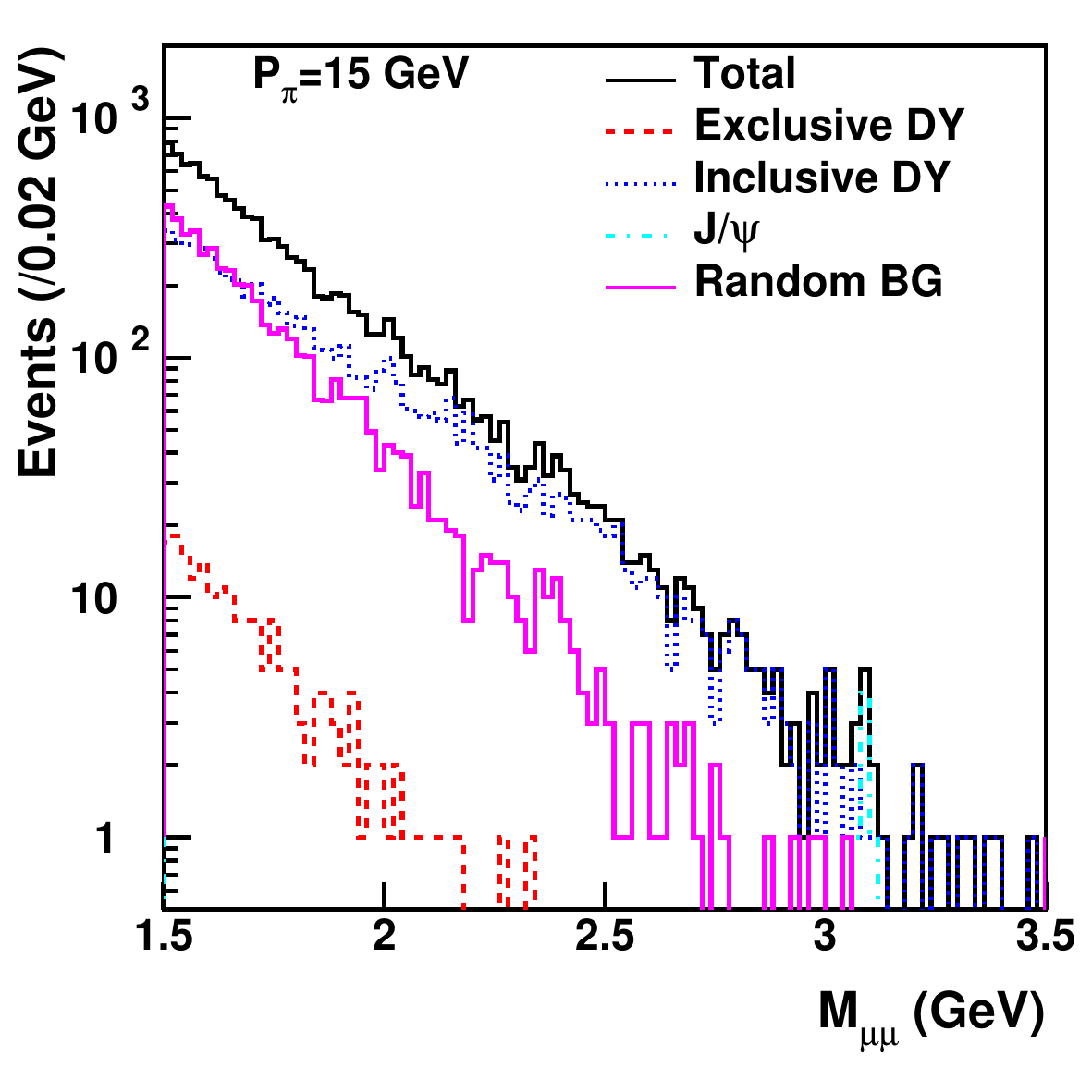}
\label{fig:invm_2}}
\hspace{-0.5cm}
\subfigure[]
{\includegraphics[width=0.33\textwidth]{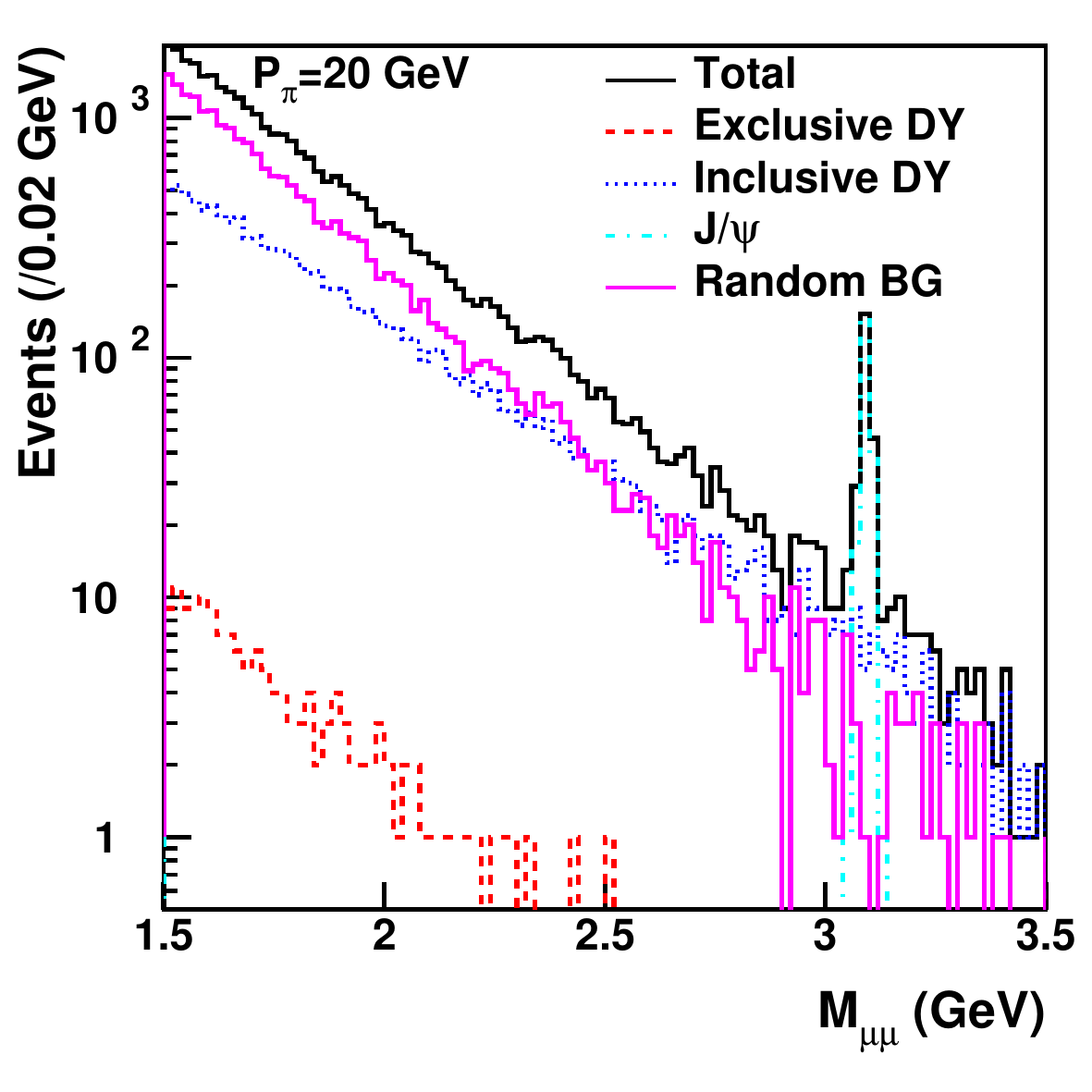}
\label{fig:invm_3}}
\caption{The Monte Carlo simulated invariant mass $M_{\mu^{+}\mu^{-}}$
  spectra of the $\mu^{+}\mu^{-}$ events with $M_{\mu^{+}\mu^{-}} >
  1.5$ GeV and $|t-t_{0}|<0.5$ GeV$^2$ for $P_{\pi}$= 10 (a), 15 (b), and 20 (c)
  GeV. Lines with different colors denote the contributions from
  various sources. The GK2013 GPDs is used for the evaluation of
  exclusive Drell-Yan process.}
\label{fig:invm}
\end{figure}

\begin{figure}[hbtp]
\centering
\hspace{-0.5cm}
\subfigure[]
{\includegraphics[width=0.33\textwidth]{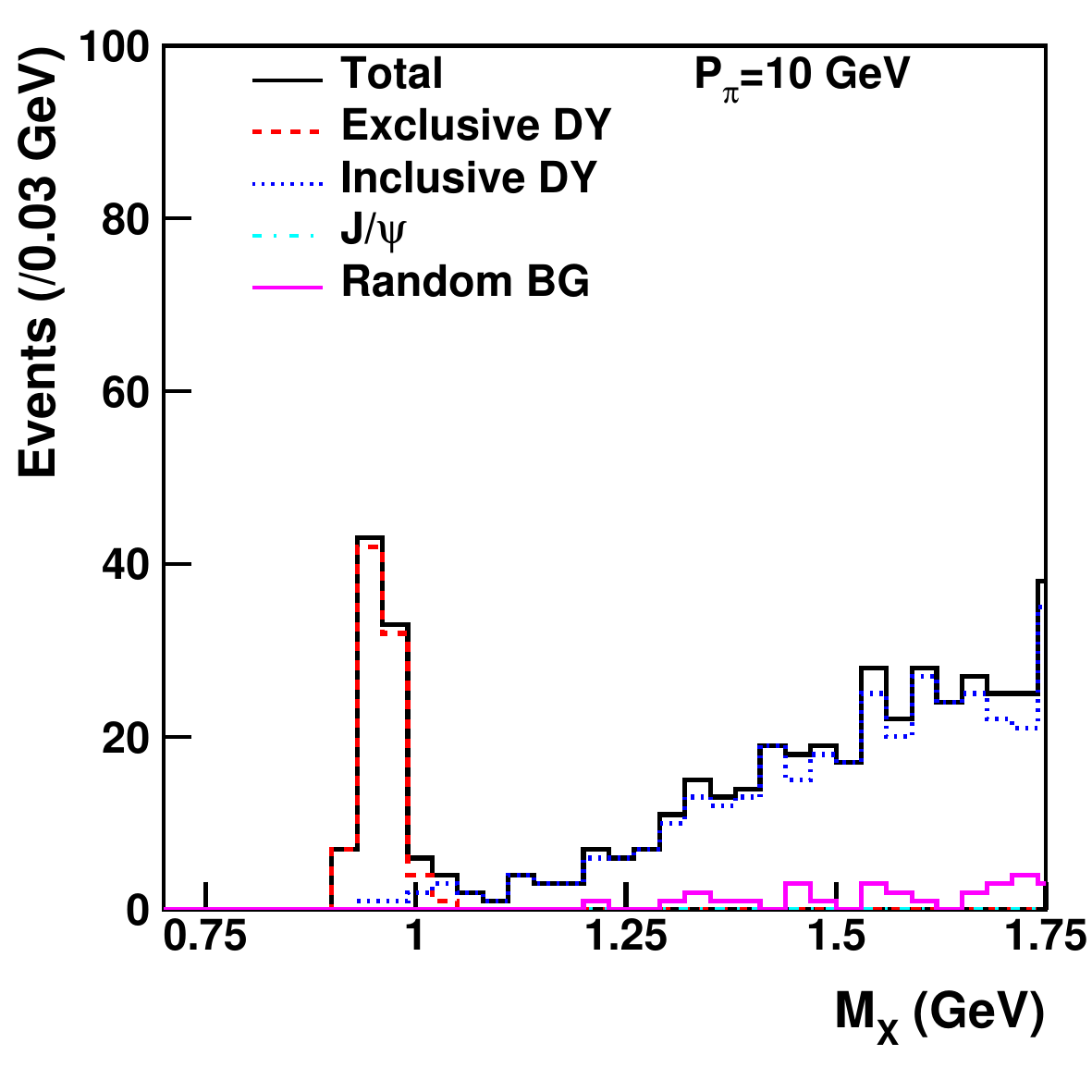}
\label{fig:mmass_1}}
\hspace{-0.5cm}
\subfigure[]
{\includegraphics[width=0.33\textwidth]{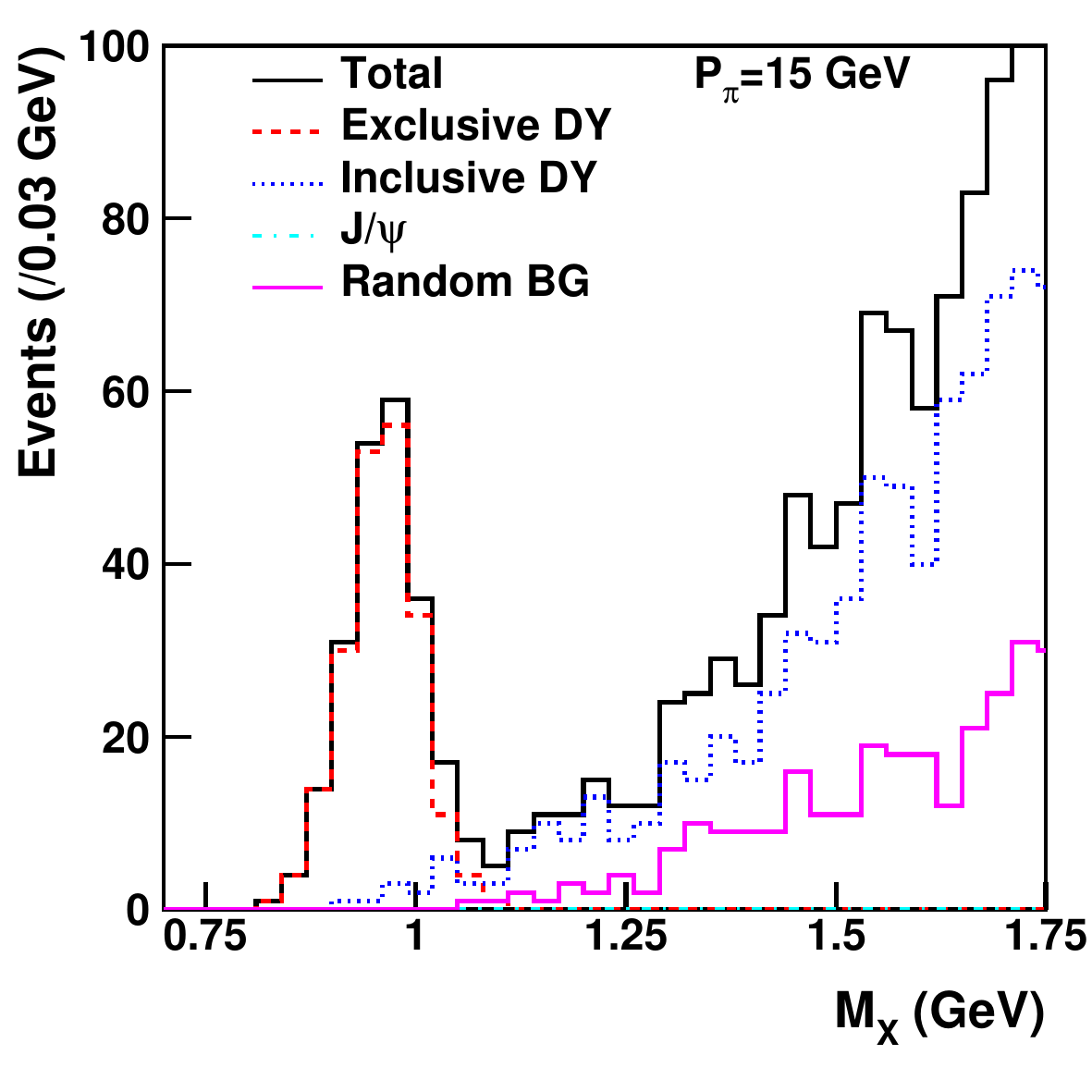}
\label{fig:mmass_2}}
\hspace{-0.5cm}
\subfigure[]
{\includegraphics[width=0.33\textwidth]{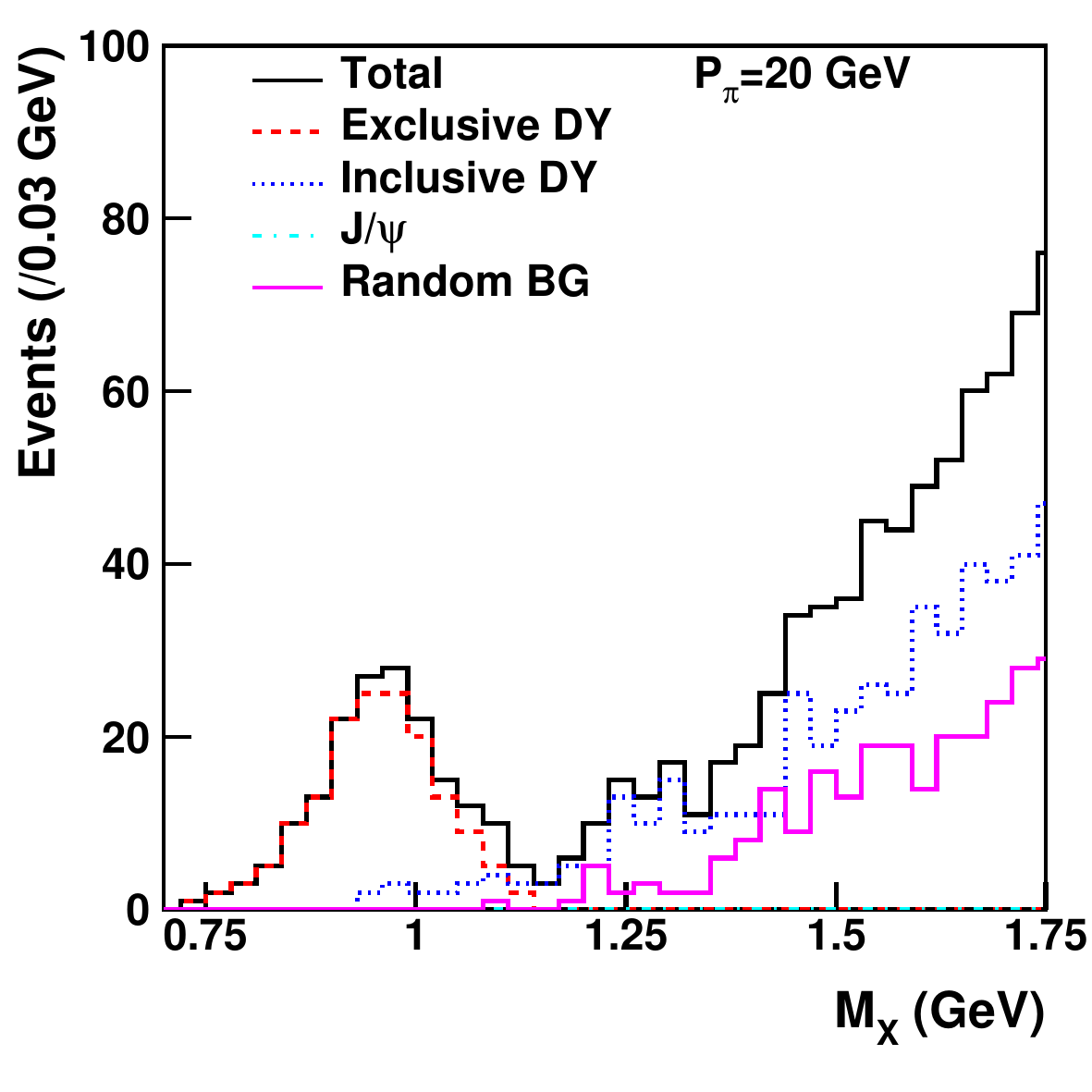}
\label{fig:mmass_3}}
\caption{The Monte Carlo simulated missing-mass $M_{X}$ spectra of the
  $\mu^{+}\mu^{-}$ events with $M_{\mu^{+}\mu^{-}} > 1.5$ GeV and
  $|t-t_{0}|<0.5$ GeV$^2$ for $P_{\pi}$= 10 (a), 15 (b), and 20 (c) GeV. Lines with
  different colors denote the contributions from various sources. The
  GK2013 GPDs is used for the evaluation of exclusive Drell-Yan
  process.}
\label{fig:mmass}
\end{figure}

Using GK2013 GPDs for the exclusive Drell-Yan process, the Monte Carlo
simulated invariant mass $M_{\mu^{+}\mu^{-}}$ and missing-mass $M_{X}$
spectra of the $\mu^{+}\mu^{-}$ events with $M_{\mu^{+}\mu^{-}} > 1.5$
GeV and $|t-t_{0}|<0.5$ GeV$^2$ for $P_{\pi}$=10, 15, and 20 GeV are
shown in Figs.~\ref{fig:invm} and ~\ref{fig:mmass}. Lines with
different colors denote the contributions from various sources:
exclusive Drell-Yan (red), inclusive Drell-Yan (blue), $J/\psi$
(green), and random background (purple), respectively. Signals of
$J/\psi$ are only visible in the invariant mass distributions for
$P_{\pi}$=15 and 20 GeV.

Figure~\ref{fig:mmass} clearly shows that the exclusive Drell-Yan events
could be identified by the signature peak at the nucleon mass in the
missing-mass spectrum for all three pion beam momenta. The $Q'$ range
of the accepted Drell-Yan events is about 1.5$-$2.5 GeV. For the case
of the lowest pion momentum $P_{\pi}$= 10 GeV, the momentum and the
missing-mass resolution is best because of the relatively low momenta
of produced muons. However the statistics of accepted $\mu^{+}\mu^{-}$
events is least due to the threshold momentum for the muon to
penetrate through the $\mu$ID system.

\begin{figure}[hbtp]
\centering
\hspace{-0.5cm}
\subfigure[]
{\includegraphics[width=0.33\textwidth]{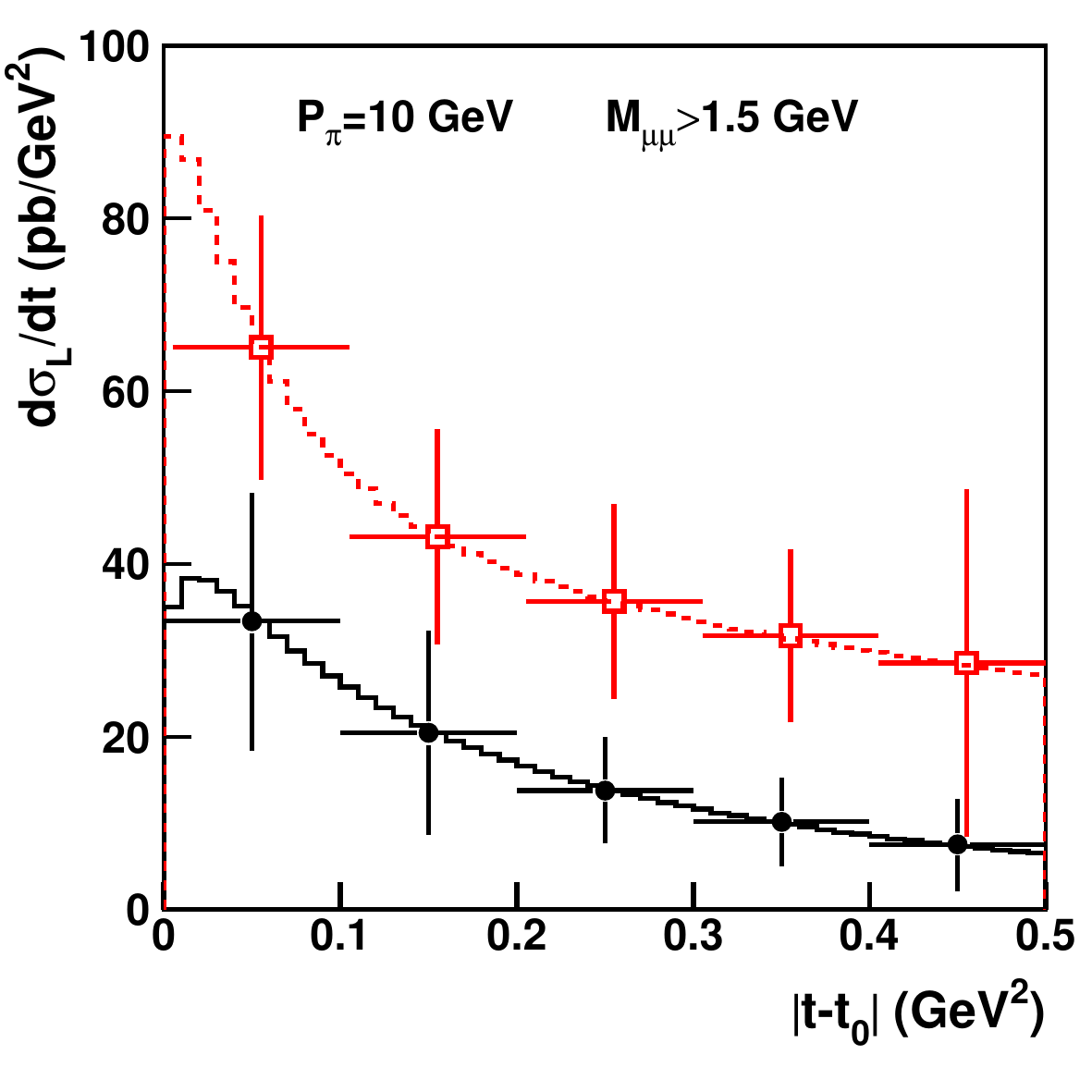}
\label{fig:edy_stat_1}}
\hspace{-0.5cm}
\subfigure[]
{\includegraphics[width=0.33\textwidth]{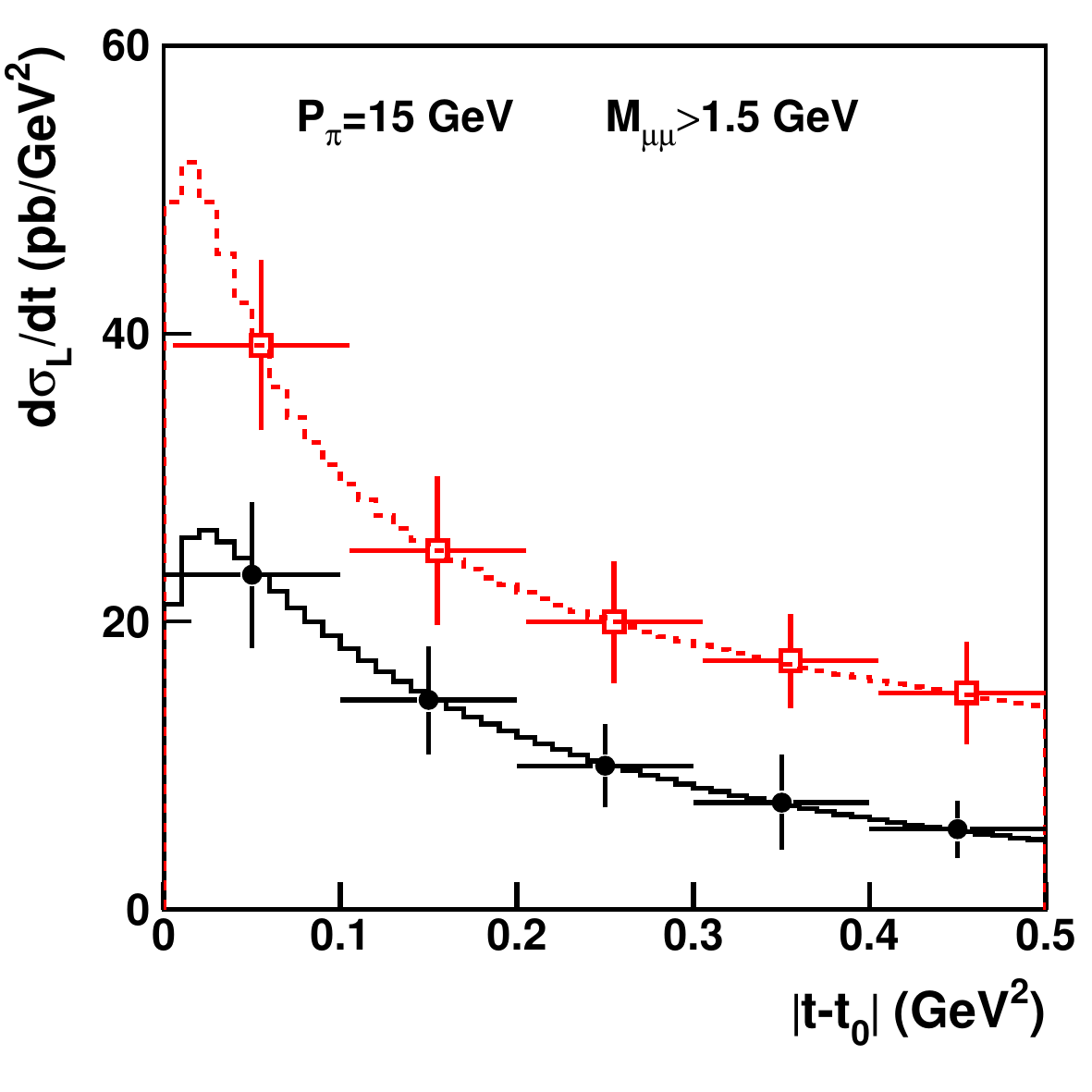}
\label{fig:edy_stat_2}}
\hspace{-0.5cm}
\subfigure[]
{\includegraphics[width=0.33\textwidth]{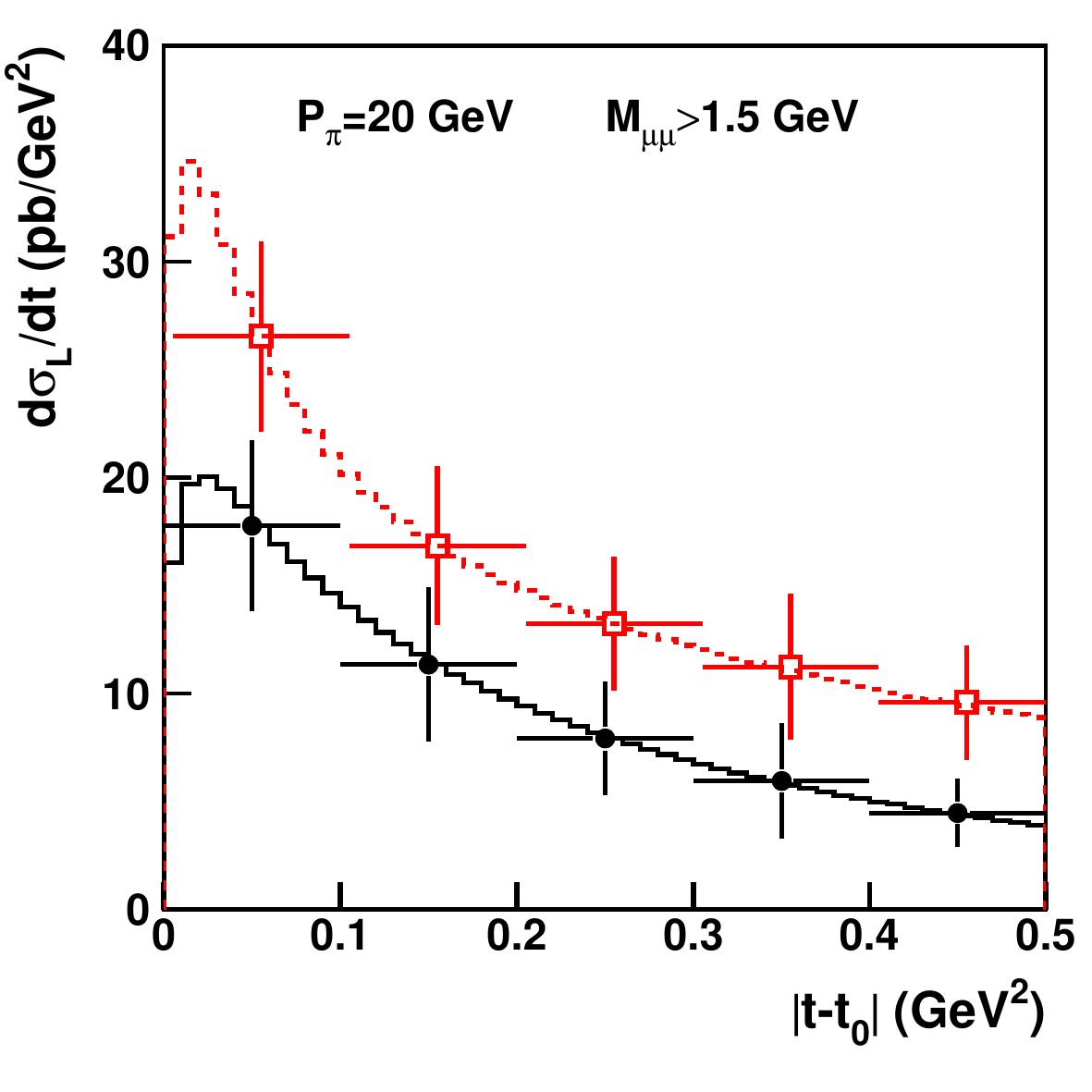}
\label{fig:edy_stat_3}}
\caption{The expected statistical errors of the exclusive Drell-Yan
  measurement for two GPDs inputs, BMP2001 (black) and GK2013 (red),
  as a function of $|t-t_{0}|$ in the dimuon mass region of
  $M_{\mu^{+}\mu^{-}} > 1.5$ GeV for 10 (a), 15 (b), and 20 (c) GeV beam
  momentum.}
\label{fig:x_sect_t}
\end{figure}

\begin{figure}[hbtp]
\centering
\includegraphics[width=0.95\textwidth]{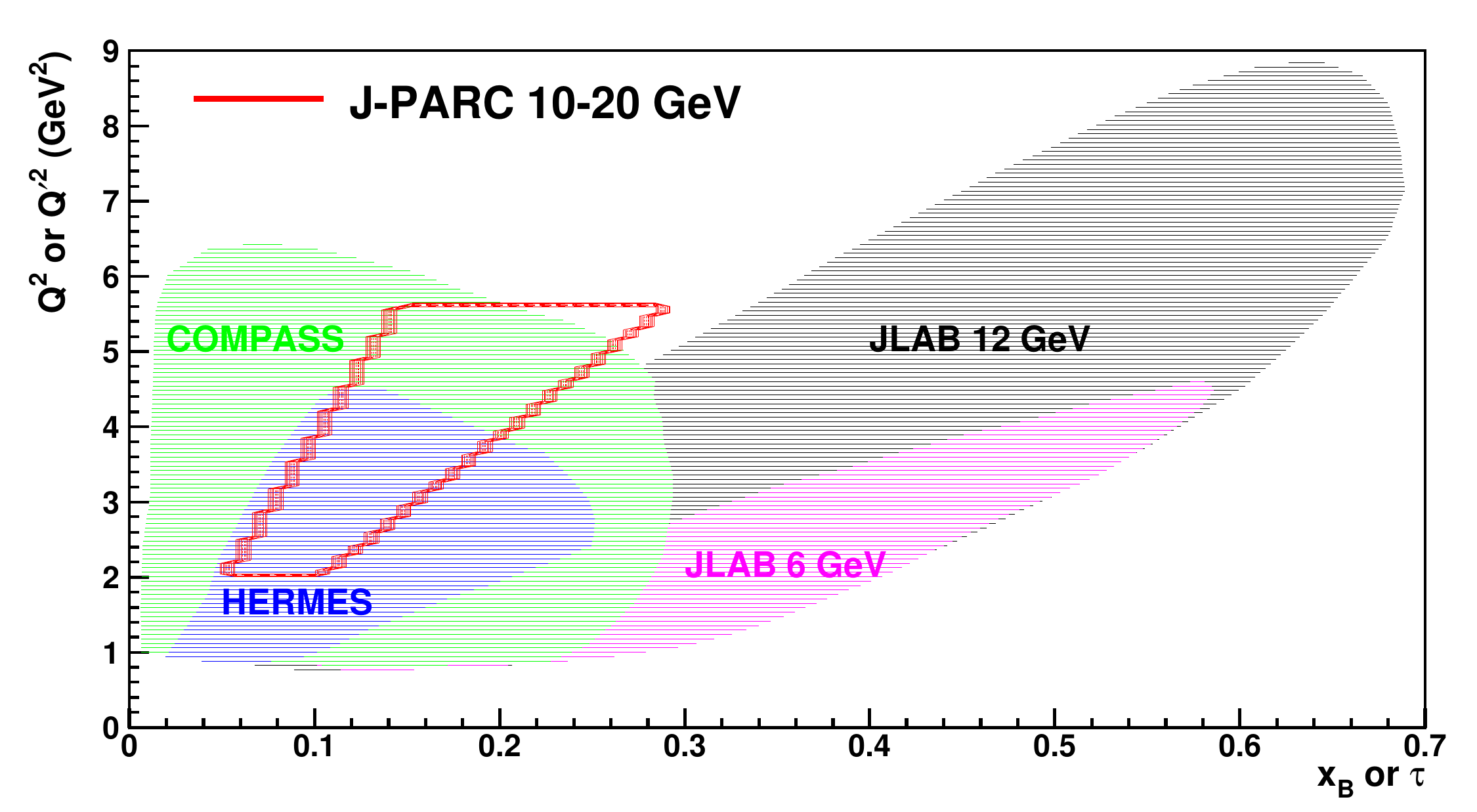}
\caption{The kinematic regions of GPDs explored by the experiments at
  JLab, HERMES and COMPASS and J-PARC (exclusive Drell-Yan). The
  region is either [$Q^2$, $x_B$] for spacelike processes or [$Q'^2$,
    $\tau$] for timelike ones.}
\label{fig:GPD_map}
\end{figure}

In Fig.~\ref{fig:x_sect_t} we show the expected statistical errors of
exclusive Drell-Yan cross sections as a function of $|t-t_{0}|$. Under
the current setting, the measurement with 15-GeV pion beam momentum is
most feasible where the GPD modeling of BMP2001 and GK2013
could be differentiated by the experiment. We compare the kinematic
regions of $Q^2$ versus $x_B$ for spacelike processes and those of
$Q'^2$ versus $\tau$ for timelike ones explored by the existing and
coming experiments in Fig.~\ref{fig:GPD_map}. Testing the universality
of nucleon GPDs through both the measurements of spacelike and
timelike processes on the same kinematic regions shall be very
important.

%%%%%%%%%%%%%%%%%%%%%%%%%%%%%%%%%%%%%%
\section{Conclusions}
\label{sec:summary}
%%%%%%%%%%%%%%%%%%%%%%%%%%%%%%%%%%%%%%

In the framework of the J-PARC E50 experiment, we addressed the
feasibility of measuring the exclusive pion-induced Drell-Yan process
in the coming high-momentum beam line of J-PARC. Detailed simulations
on signal reconstruction efficiency as well as on rejection of the
most severe random background channel were performed for the pion beam
momentum in the range of 10$-$20 GeV. A clean signal of exclusive
pion-induced Drell-Yan process can be identified in the missing-mass
spectrum of dimuon events with 2$-$4 fb$^{-1}$ integrated
luminosity. The statistics accuracy is adequate for discriminating
between the predictions from two current GPD modelings. The
realization of this measurement will represent not only a new approach
of accessing nucleon GPDs and pion DAs in the timelike process, but also a
novel test of the factorization of an exclusive Drell-Yan process
associated with timelike virtuality and the universality of GPDs in
spacelike and timelike processes.  Since both the inclusive and
exclusive Drell-Yan events could be measured simultaneously, the data
could reveal interesting features in the transition from inclusive
Drell-Yan to the semi-exclusive and exclusive limits~\cite{Teryaev}.

The pion pole in the GPD $\tilde{E}$ is expected to give a dominant
contribution in the cross sections at small $|t|$. The cross sections
at this pion-pole dominance region will provide a unique opportunity
to access the pion timelike form factor other than the approach of
$e^+e^-$ annihilation process. The input cross section used for the
exclusive Drell-Yan events in this work is the prediction within the
factorization approach using the partonic hard scattering in the
leading order in $\alpha_s$ convoluted with the leading-twist pion DAs
and nucleon GPDs. We call for further theoretical progress which might
improve the prediction on production cross sections, e.g., the work of
Ref.~\cite{Goloskokov:2015zsa}.

\section*{ACKNOWLEDGMENTS}
We acknowledge helpful discussions with Hiroyuki Kawamura, Peter
Kroll, Hsiang-nan Li, Bernard Pire, Kotaro Shirotori, and Masashi
Wakamatsu. This work was supported in part by the Ministry of Science
and Technology of Taiwan, the Ministry of Education, Culture, Sports,
Science and Technology of Japan (Grants No. 25105010, No. 25610058,
and No. 26287040) and the U.S. National Science Foundation.

%%%%%%%%%%%%%%%%%%%%%%%%%%%%%%%%%%%%%%%%%%%%%%%%%%%%%%%%%%%%%%%%%%%%%%%%

\end{document}